\documentclass[final,5p,times,twocolumn]{elsarticle}
\usepackage{microtype}
\usepackage[T1]{fontenc}
\usepackage{amsmath,amssymb,amsfonts}
\usepackage{bm}

\usepackage{algorithm}
\usepackage{algorithmic}
\usepackage{graphicx}
\usepackage{textcomp}
\usepackage{url}
\usepackage{xurl}
\usepackage[caption=false,font=normalsize,labelfont=sf,textfont=sf]{subfig}
\usepackage{stfloats}
\usepackage{verbatim}
\usepackage{booktabs}
\usepackage{hhline}
\usepackage[dvipsnames,table]{xcolor}
\usepackage{pifont}
\usepackage{wrapfig}
\usepackage[normalem]{ulem}
\usepackage{colortbl}
\usepackage{etoolbox}
\usepackage{array,tabularx}
\usepackage{placeins}
\usepackage{enumitem}
\usepackage{adjustbox}
\usepackage{makecell}
\usepackage{multirow}
\usepackage{ragged2e}
\usepackage{needspace}

\usepackage[colorlinks=true,linkcolor=blue,citecolor=blue,urlcolor=blue]{hyperref}

\biboptions{sort&compress}

\definecolor{LightGray}{gray}{0.92}
\definecolor{TIISPureBlue}{RGB}{0,92,175}
\newcommand{\cmark}{\textcolor{ForestGreen}{\ding{51}}}

\newcommand{\chk}{\scriptsize\cmark}

\newcolumntype{Y}{>{\centering\arraybackslash}X}
\newcolumntype{C}[1]{>{\centering\arraybackslash}m{#1}}
\newcolumntype{L}[1]{>{\centering\arraybackslash}m{#1}}
\newcolumntype{R}[1]{>{\centering\arraybackslash}m{#1}}
\newcolumntype{P}[1]{>{\RaggedRight\arraybackslash}p{#1}}
\usepackage{array}
\usepackage{graphicx}
\usepackage{multirow}
\usepackage[table]{xcolor}
\usepackage{amssymb}

\newcolumntype{L}[1]{>{\raggedright\arraybackslash}p{#1}}
\newcolumntype{C}[1]{>{\centering\arraybackslash}p{#1}}
\newcommand{\rotcite}[1]{\rotatebox[origin=c]{90}{\strut\cite{#1}}}
\newcommand{\rottext}[1]{\rotatebox[origin=c]{90}{\strut #1}}
\DeclareRobustCommand{\CrossRefArticle}[1]{%
  \textcolor{blue}{\uline{\href{#1}{Article (CrossRef Link)}}}%
}
\newcommand{\figref}[1]{\hyperref[#1]{\textcolor{TIISPureBlue}{Fig.~\ref*{#1}}}}
\newcommand{\tabref}[1]{\hyperref[#1]{\textcolor{TIISPureBlue}{Table~\ref*{#1}}}}

\setlist[itemize]{topsep=2pt,itemsep=1pt,parsep=0pt,partopsep=0pt,leftmargin=*}

\journal{Journal name}

\begin{document}

\begin{frontmatter}

\title{Robustness Evaluation and Detection of Transferable Adversarial Attacks in ML-Based NIDS}

\author[inst1]{Huda Ali Alatawi\corref{cor1}}
\ead{haladheedi@ut.edu.sa}
\cortext[cor1]{Corresponding author}

\affiliation[inst1]{organization={Department of Information Technology, College of Computer Science and Information Technology, University of Tabuk},
            city={Tabuk},
            country={Saudi Arabia}}

\begin{abstract}
Machine learning-based network intrusion detection systems (ML-based NIDS) are vulnerable to adversarial evasion, where malicious samples are perturbed to evade detection and be misclassified as benign. Despite growing research on adversarial attacks and defenses for ML-based NIDS, comparative evaluations of multiple attack types, detection models, and defense strategies under a common setting remain limited. In this paper, we evaluate eight adversarial evasion attacks, fifteen detection models, and three representative defense strategies using the NF-UQ-NIDS dataset, which includes recent traditional and IoT network traffic with twenty distinct attack categories. The evaluation compares model performance on clean test data and on robustness evaluation sets that include adversarial samples, analyzes attack success consistency across models, and examines the effect of defense strategies on adversarial robustness. To support model comparison, we introduce the Robustness Index (RI), a compact comparative metric that rewards high balanced accuracy and macro-F1 score computed on the robustness evaluation set while penalizing high attack success rate (ASR). We further present AR-NIDS, a two-stage framework that uses an adversarially trained ensemble to distinguish normal, attack, and adversarial samples, followed by an adversarial attack classifier to identify the attack type. Under the evaluated transfer-based setting, the proposed adversarially trained ensemble achieves the strongest overall trade-off between classification performance on the robustness evaluation set and evasion resistance, reducing the average ASR from 0.41 to 0.03 and achieving an RI of 0.98.
\end{abstract}

\begin{keyword}
Network Intrusion Detection \sep
Adversarial Defense \sep
Adversarial Robustness \sep
Adversarial Examples \sep
Adversarial Attacks \sep
Adversarial Machine Learning \sep
Network Security
\end{keyword}

\end{frontmatter}

%\clearpage

%%%%%%%%%%%%%%%%%%%%%%%%%%%%%%%%%%%%%%%%%%%%%%%%%%%%%%%%%%%%%%%%%%%%%

\section{Introduction}\label{introduction}

Network Intrusion Detection Systems (NIDS) protect networks by monitoring traffic for signs of security violations. Signature-based NIDS compare traffic patterns to a database of known attack signatures but are ineffective against zero-day attacks~\cite{buczak2015survey}. Anomaly-based NIDS, on the other hand, use machine learning to detect deviations from normal traffic behavior, enabling them to identify previously unseen attacks and enhance network security~\cite{buczak2015survey,ahmad2021network}. Despite these advantages, ML-based NIDS are vulnerable to adversarial examples, where malicious samples are intentionally perturbed to deceive machine learning models and evade detection~\cite{he2023adversarial}. These adversarial examples exploit model vulnerabilities and may transfer across models, meaning that samples generated against one model can also fool other detection models~\cite{szegedy2013intriguing}. This transferability is particularly important in NIDS because an attacker may use a surrogate IDS model to generate adversarial samples and then attempt to evade a different deployed detector without direct access to its internal parameters~\cite{he2023adversarial}.

Significant research has been conducted on evaluating and defending against adversarial attacks in ML-based NIDS~\cite{rigaki2017adversarial,debicha12detect, rashid2022adversarial, martins2019analyzing, debicha2021adversarial, fu2021robust, piplai2020nattack, schneider2021evaluating, abdelaty2021gadot, zhang2020tiki}. Despite these advancements, existing studies often examine isolated aspects of the problem, such as a limited set of attacks, a small number of models, or a single defense mechanism. Comparative assessments that jointly compare attack behavior, model robustness, and defense effectiveness under a common evaluation setting remain limited. This issue is further complicated by the continued use of older datasets in parts of the literature, which may reduce the relevance of findings to recent network environments~\cite{ring2019survey}. Therefore, evaluating multiple adversarial attacks, detection models, and defense strategies using recent NetFlow-based data is important for understanding the robustness of ML-based NIDS under modern adversarial conditions.

In ML-based NIDS, robustness refers to a model's ability to maintain reliable detection performance when facing adversarial attacks or other perturbations~\cite{wang2023robustness,schneider2021evaluating}. Robustness evaluations often report detection performance and attack success rate (ASR) separately. While these metrics are useful, reporting them in isolation can obscure important trade-offs. For example, a model may achieve strong performance on clean data while remaining vulnerable to adversarial evasion, or it may reduce ASR while suffering degraded classification performance. This motivates the need for a compact comparative metric that considers both classification performance when adversarial samples are included and resistance to evasion when comparing ML-based NIDS models.

Research on enhancing ML-NIDS robustness has explored strategies such as adversarial training~\cite{rashid2022adversarial, martins2019analyzing, fu2021robust, piplai2020nattack, schneider2021evaluating, abdelaty2021gadot, zhang2020tiki}, ensemble learning~\cite{rigaki2017adversarial,debicha12detect}, and specialized adversarial detectors~\cite{debicha2021adversarial}. These approaches have shown improvements in some settings, but their effectiveness can vary depending on the attack type, model architecture, and evaluation scenario. Ensemble methods may improve predictive performance on clean data but do not necessarily guarantee adversarial robustness. Similarly, adversarial training may reduce attack success but can be sensitive to the types of adversarial examples used during training. These trade-offs highlight the need to compare representative defense strategies under the same transfer-based adversarial setting.

In this paper, we present an empirical robustness evaluation of ML-based NIDS under transferable adversarial attacks. We use the NF-UQ-NIDS dataset~\cite{sarhan2021netflow}, which contains recent traditional and IoT network traffic with twenty attack categories. The study evaluates eight adversarial evasion attacks, fifteen detection models, and three representative defense strategies. The adversarial examples are generated against a feed-forward deep neural network surrogate model and evaluated across other ML-based NIDS models to assess transferability. This evaluation setting allows us to analyze how different models and defenses respond to adversarial samples generated under a common threat model.

\textbf{Contributions:} The contributions of this paper are as follows:
\begin{itemize}
\item We conduct a comparative evaluation of adversarial attack performance, model robustness, and defense effectiveness. Our assessment includes eight attack types, fifteen detection models, and three representative defense strategies using the NF-UQ-NIDS dataset~\cite{sarhan2021netflow}, which includes recent traditional and IoT network traffic with twenty attack categories. We compare models' performance on clean test data and robustness evaluation sets (Sections~\ref{Models Performance Over Clean Test Data} and~\ref{Models Performance: Clean vs. Adversarial}), analyze attacks' performance consistency across detection models (Section~\ref{Adversarial Attacks Performance}), and examine the effect of defense strategies on adversarial attack success (Section~\ref{Defense Effectiveness Against Adversarial Attacks}).

\item We introduce the Robustness Index (RI), a compact comparative metric that combines balanced accuracy and macro-F1 score computed on the robustness evaluation set with attack success rate. RI supports model comparison by capturing the trade-off between classification performance when adversarial samples are included and resistance to adversarial evasion (Section~\ref{Robustness Index}).

\item We introduce the Adversarial-Robust NIDS (AR-NIDS), a two-stage framework that uses an adversarially trained ensemble to distinguish normal, attack, and adversarial samples, followed by an adversarial attack classifier to identify the attack type (Section~\ref{Adversarial-Robust Network Intrusion Detection System (AR-NIDS)}). Under the evaluated transfer-based setting, the adversarially trained ensemble improves empirical robustness compared with the evaluated baseline and defense models (Section~\ref{Evaluation of Adversarial-Robust NIDS}).
\end{itemize}

\begin{comment}
The remainder of this paper is structured as follows. Section~\ref{Background} provides background on adversarial attacks and ML-based NIDS performance metrics. Section~\ref{Related Work} reviews related work on adversarial attack defense. Section~\ref{Framework and Index for ML-based NIDS Robustness} introduces the RI and AR-NIDS framework. Section~\ref{experimental setup} describes the experimental setup. Section~\ref{Robustness Evaluation of ML-based NIDS to Adversarial Attacks} evaluates baseline models on clean test data and robustness evaluation sets, measures their robustness using RI, and analyzes attack performance consistency. Section~\ref{Evaluation of Adversarial-Robust NIDS} evaluates the AR-NIDS framework and compares the evaluated defense strategies. Section~\ref{discussion} discusses the implications and limitations of the results. Finally, Section~\ref{conclusion} concludes the paper.
\end{comment}

%=========================================================%
\section{Background}\label{Background}

This section provides background on adversarial attacks and the evaluation metrics used in this study. We first summarize the main categories of adversarial attacks and then describe the evasion attacks considered in our evaluation. We then introduce the classification metrics used to assess ML-based NIDS performance and to formulate the proposed Robustness Index (RI).

%%%%%%%%%%%%%%%%%%%%%%%%%%%%%%%%%%%%%%%%%%%%%%%%%%%%%%%%%%%%%%%%%%%%%%%%%%%%%%%%
\subsection{Adversarial Attacks}\label{Adversarial Attacks}

Adversarial examples (AEs) are inputs or input representations that are intentionally perturbed to cause incorrect predictions by a machine learning model~\cite{chakraborty2021survey,kurakin2016adversarial}. In general, adversarial example generation aims to find a perturbed sample $x^*$ that remains close to the original sample $x$ while changing the model's prediction toward an adversarial objective~\cite{biggio2018wild,chakraborty2021survey,kurakin2016adversarial}. This can be expressed as finding a perturbation $\gamma$ such that $x^*=x+\gamma$, where $\gamma$ is constrained to be small according to a selected distance or perturbation measure while still inducing misclassification.

Adversarial attacks can be broadly categorized according to the attacker's goal and stage of intervention. Poisoning attacks target the training process by injecting malicious or manipulated samples into the training data to degrade the learned model. Evasion attacks target the inference stage by modifying test samples so that malicious instances are misclassified. Black-box or query-based attacks assume limited knowledge of the target model and may rely on model queries, substitute models, or transferability to generate adversarial samples~\cite{chakraborty2021survey}. This paper focuses on evasion attacks, where malicious samples are perturbed to evade detection by ML-based NIDS models.

The attacks evaluated in this study include gradient-based, optimization-based, and query-based methods. The Jacobian-based Saliency Map Attack (JSMA) identifies influential input features using the Jacobian matrix of the model output and perturbs selected features to induce misclassification~\cite{papernot2016limitations}. The Carlini-Wagner (CW) attack formulates adversarial example generation as an optimization problem that searches for small perturbations capable of changing the model decision under a selected distance metric~\cite{carlini2017towards}. DeepFool iteratively estimates the minimum perturbation required to cross the model's decision boundary, producing adversarial examples with small changes relative to the original input~\cite{moosavi2016DeepFool}. The Fast Gradient Sign Method (FGSM) is a one-step gradient-based attack that perturbs the input in the direction that increases the model loss~\cite{goodfellow2014explaining}. Projected Gradient Descent (PGD) extends this idea by applying iterative gradient-based perturbations and projecting the perturbed sample back into an allowed perturbation region~\cite{madry2017towards}. Zeroth Order Optimization (ZOO) is a black-box optimization attack that estimates gradients numerically through model queries, allowing adversarial examples to be generated without direct access to model gradients~\cite{chen2017zoo}. HopSkipJump (HSJ) is a decision-based black-box attack that uses model output decisions to search for boundary-crossing perturbations with limited information about the target model~\cite{chen2020hopskipjumpattack}.

%%%%%%%%%%%%%%%%%%%%%%%%%%%%%%%%%%%%%%%%%%%%%%%%%%%%%%%%%%%%%%%%%%%%%%%%%%%%%%%%
\subsection{Evaluation Metrics for ML-based NIDS}\label{Evaluation Metrics for ML-based NIDS}

Binary classification in ML-based NIDS commonly categorizes network traffic into two classes: attack and normal. In this setting, attack traffic is treated as the positive class and normal traffic as the negative class. The performance of a classifier can be summarized using a confusion matrix, which reports correct and incorrect predictions for both classes.

\textbf{Confusion Matrix:} The confusion matrix summarizes the relationship between actual and predicted classes:
\[
    \begin{array}{|c|c|c|}
    \hline
    & \textbf{Predicted: Attack} & \textbf{Predicted: Normal} \\
    \hline
    \textbf{Actual: Attack} & \text{TP (True Positive)} & \text{FN (False Negative)} \\
    \textbf{Actual: Normal} & \text{FP (False Positive)} & \text{TN (True Negative)} \\
    \hline
    \end{array}
\]

where:
\begin{itemize}
    \item \textbf{TP:} attack samples correctly classified as attacks.
    \item \textbf{FP:} normal samples incorrectly classified as attacks.
    \item \textbf{TN:} normal samples correctly classified as normal.
    \item \textbf{FN:} attack samples incorrectly classified as normal.
\end{itemize}

The following metrics are used to evaluate detection performance:

\begin{itemize}
    \item \textbf{Accuracy:} The proportion of correctly classified samples among all evaluated samples.
    \begin{equation}
    \text{Accuracy} = \frac{TP + TN}{TP + TN + FP + FN}
    \end{equation}

    \item \textbf{Precision:} The proportion of samples predicted as attacks that are truly attacks. High precision indicates fewer false alarms.
    \begin{equation}
    \text{Precision} = \frac{TP}{TP + FP}
    \end{equation}

    \item \textbf{Recall:} The proportion of actual attack samples that are correctly detected. High recall indicates that fewer attacks are missed.
    \begin{equation}
    \text{Recall} = \frac{TP}{TP + FN}
    \end{equation}

    \item \textbf{F1-score:} The harmonic mean of precision and recall. It provides a balanced measure when both false positives and false negatives are important.
    \begin{equation}
    \text{F1-score} = 2 \times \frac{\text{Precision} \times \text{Recall}}{\text{Precision} + \text{Recall}}
    \end{equation}

    \item \textbf{Balanced Accuracy:} The average of the true positive rate and true negative rate. It is useful for imbalanced datasets because it gives equal importance to attack and normal classes.
    \begin{equation}
    \text{Balanced Accuracy} = \frac{1}{2} \left( \frac{TP}{TP + FN} + \frac{TN}{TN + FP} \right)
    \end{equation}

    \item \textbf{Attack Success Rate (ASR):} The proportion of adversarial attack samples that successfully evade detection. In this study, an adversarial attack sample is considered successful when it is misclassified as normal.
    \begin{equation}
    \text{ASR} = \frac{\text{Number of successful adversarial samples}}{\text{Total number of adversarial samples}}
    \end{equation}
\end{itemize}

These metrics capture different aspects of ML-based NIDS performance. Accuracy reflects overall correctness, but it may be misleading under class imbalance. Precision and recall focus on false alarms and missed attacks, respectively. F1-score balances precision and recall, while balanced accuracy accounts for both attack and normal classes. ASR directly measures adversarial evasion success. These metrics form the basis for the robustness evaluation and for the Robustness Index introduced in Section~\ref{Robustness Index}.
%%%%%%%%%%%%%%%%%%%%%%%%%%%%%%%%%%%%%%%%%%%%%%%%%%%%%%%%%

\section{Related Work}~\label{Related Work}

This section reviews related work on adversarial defenses for ML-based NIDS, with a focus on adversarial training, ensemble learning, and comparative robustness evaluation. The purpose of this review is to position the present study within existing efforts and to clarify the gaps addressed by our empirical evaluation.

\subsection{Adversarial Training (AT)}

Adversarial training is one of the most widely used defense strategies against adversarial evasion. It improves model robustness by augmenting the training process with adversarial examples, allowing the model to learn decision boundaries that are less sensitive to adversarial perturbations. In ML-based NIDS, adversarial training has been explored across different models, datasets, and attack settings.

Rashid et al.~\cite{rashid2022adversarial} evaluated the impact of five adversarial attacks on a DNN-based cyberattack detection model and investigated adversarial training as a defense mechanism. Martins et al.~\cite{martins2019analyzing} assessed six NIDS classifiers against four adversarial attacks and reported that adversarial training improved robustness in their evaluated setting. Debicha et al.~\cite{debicha2021adversarial} studied FGSM and PGD attacks against a DNN-based intrusion detection model and showed that adversarial training can reduce the impact of these attacks. Similarly, Fu et al.~\cite{fu2021robust} found that adversarial training improved the robustness of LSTM, CNN, and GRU models against FGSM-based adversarial examples.

Other studies have reported more nuanced findings. Piplai et al.~\cite{piplai2020nattack} observed that adversarial training was not highly effective for GAN-based models under FGSM attacks. Schneider et al.~\cite{schneider2021evaluating} evaluated the robustness of several classical ML models, including DT, RF, LR, LDA, QDA, and MLP, and tested defense strategies such as mutable feature removal and ensemble-based methods. Abdelaty et al.~\cite{abdelaty2021gadot} proposed a GAN-based approach for adversarial training, while Zhang et al.~\cite{zhang2020tiki} examined multiple defense approaches, including adversarial training, query detection, and model voting ensembling.

Overall, prior work suggests that adversarial training can improve robustness in ML-based NIDS, but its effectiveness depends on the attack type, model architecture, training data, and evaluation setting. In particular, adversarial training may improve robustness against attacks similar to those seen during training, but its effectiveness against diverse or transferable attacks remains an important evaluation question.

\subsection{Ensemble Learning (EL)}

Ensemble learning combines multiple models to improve predictive performance and reduce dependence on a single classifier. In intrusion detection, ensemble methods are often used to improve accuracy, stability, and generalization by aggregating predictions from diverse base learners. From an adversarial robustness perspective, ensembles may reduce the risk that an attacker can exploit the weakness of a single model. However, ensemble learning does not automatically guarantee adversarial robustness, especially when different models share similar decision-boundary vulnerabilities.

Rigaki and Garcia~\cite{rigaki2017adversarial} evaluated DT, SVM, RF, and voting ensemble models against FGSM and JSMA attacks. Their results showed that ensemble models did not necessarily outperform individual models under adversarial conditions. In contrast, Debicha et al.~\cite{debicha12detect} studied FGSM and PGD attacks on SVM, DT, LR, RF, and LDA models and reported that ensemble models showed improved robustness in their setting.

These mixed findings indicate that the effectiveness of ensemble learning as an adversarial defense is context-dependent. Ensemble diversity, base-model selection, attack type, and evaluation protocol all influence whether an ensemble improves or weakens robustness. Therefore, ensemble learning should be evaluated alongside other defense mechanisms rather than assumed to be inherently robust.

Table~\ref{tab:attacks-defenses-comparison} summarizes related studies according to the adversarial attacks and defense mechanisms considered. The table highlights the breadth of prior work on adversarial attacks against NIDS and shows that adversarial training and ensemble learning are commonly studied defense strategies. It also positions the present work as a comparative evaluation that jointly considers multiple attack types, multiple detection models, and representative defense strategies under a common transfer-based setting.

%%%%%%%%%%%%%%%%%%%%%%%%%%%%%%%%%%%%%%%%%%%%%%%%%%%%%%%%
%%%%%%%%%%%%%%%%%%%%%%%%%%%%%%%%%%%%%%%%%%%%%%%%%%%%%%%%%%%%%%%%%%%%%%%%%%%%%%%%%

%%%%%%%%%%%%%%%%%%%%%%%%%%%%%%%%%%%%%%%%%%%%%%%%%%%%%%%%%%%%%%%%%%%%%%%%%%%%%%%%%
\begin{table*}[!t]
\centering
\caption{Comparison of adversarial attacks and defense mechanisms in related ML-based NIDS studies.}
\label{tab:attacks-defenses-comparison}

\scriptsize
\setlength{\tabcolsep}{1.1pt} % was 2.2pt
\renewcommand{\arraystretch}{1.12}
\arrayrulecolor{gray!55}
\setlength{\arrayrulewidth}{0.30pt}

\resizebox{0.98\textwidth}{!}{%
\begin{tabular}{@{}|C{0.55cm}|L{1.15cm}|*{30}{C{0.38cm}|}@{}}
\hline
\rowcolor{gray!12}
\textbf{Type} & \textbf{Method}
& \rotcite{rashid2022adversarial}
& \rotcite{jadidi2022security}
& \rotcite{rigaki2017adversarial}
& \rotcite{martins2019analyzing}
& \rotcite{debicha2021adversarial}
& \rotcite{fu2021robust}
& \rotcite{piplai2020nattack}
& \rotcite{schneider2021evaluating}
& \rotcite{abdelaty2021gadot}
& \rotcite{chaitou2021assessing}
& \rotcite{shu2022omni}
& \rotcite{sauka2022adversarial}
& \rotcite{usama2019generative}
& \rotcite{zhang2020tiki}
& \rotcite{benzaid2020robust}
& \rotcite{debicha12detect}
& \rotcite{mccarthy2021feature}
& \rotcite{novaes2021adversarial}
& \rotcite{nugraha2021detecting}
& \rotcite{peng2020detecting}
& \rotcite{lin2022elat}
& \rotcite{nowroozi2022spritz}
& \rotcite{nguyen2022preventing}
& \rotcite{alahmed2022mitigation}
& \rotcite{chauhan2020polymorphic}
& \rotcite{shieh2022detection}
& \rotcite{han2020practical}
& \rotcite{usama2019generative}
& \rotcite{Abusnaina2019a}
& \rottext{\textbf{Ours}} \\
\hline

\multirow{7}{*}{\rotatebox[origin=c]{90}{\textbf{Attack}}}
& JSMA
& \chk &      & \chk & \chk &      &      &      &      &      &      & \chk &      &      &      &      &      &      &      &      &      & \chk & \chk & \chk &      &      &      &      &      &      & \chk \\ \cline{2-32}

& C\&W
& \chk &      &      & \chk &      &      &      &      &      &      & \chk &      &      &      &      &      &      &      &      &      & \chk & \chk & \chk &      &      &      &      &      & \chk & \chk \\ \cline{2-32}

& DeepFool
& \chk &      &      & \chk &      &      &      &      &      &      & \chk & \chk &      &      &      &      &      &      &      &      &      & \chk & \chk &      &      &      &      &      & \chk & \chk \\ \cline{2-32}

& FGSM
& \chk & \chk &      & \chk & \chk & \chk & \chk &      & \chk &      & \chk & \chk &      &      & \chk & \chk & \chk &      &      & \chk & \chk & \chk & \chk &      &      &      &      &      &      & \chk \\ \cline{2-32}

& PGD
&      &      &      &      & \chk &      &      & \chk &      &      &      & \chk &      &      &      & \chk &      &      &      & \chk & \chk & \chk &      &      &      &      &      &      & \chk & \chk \\ \cline{2-32}

& ZOO
&      &      &      &      &      &      &      &      &      &      &      &      &      &      &      &      &      &      &      &      & \chk &      &      & \chk &      &      &      &      &      & \chk \\ \cline{2-32}

& HSJ
&      &      &      &      &      &      &      &      &      &      &      &      &      & \chk &      &      &      &      &      &      &      &      &      &      &      &      &      &      &      & \chk \\
\hline\hline

\multirow{3}{*}{\rotatebox[origin=c]{90}{\textbf{Defense}}}
& EL
&      &      & \chk &      &      &      &      & \chk &      &      & \chk &      &      & \chk &      & \chk &      &      &      &      & \chk & \chk & \chk &      &      &      & \chk &      &      & \chk \\ \cline{2-32}

& AT
& \chk & \chk &      & \chk & \chk & \chk & \chk & \chk & \chk & \chk &      & \chk & \chk & \chk & \chk &      &      & \chk & \chk & \chk & \chk &      &      & \chk & \chk & \chk &      & \chk & \chk & \chk \\ \cline{2-32}

& ATE
&      &      &      &      &      &      &      &      &      &      &      &      &      &      &      &      &      &      &      &      &      &      &      &      &      &      &      &      &      & \chk \\
\hline
\end{tabular}%
}

\vspace{0.35em}
\begin{minipage}{0.98\textwidth}
\scriptsize
\textit{Note.} \chk indicates that the attack or defense mechanism was considered in the study.
EL: ensemble learning; AT: adversarial training; ATE: adversarially trained ensemble.
\end{minipage}

\arrayrulecolor{black}
\end{table*}
%%%%%%%%%%%%%%%%%%%%%%%%%%%%%%%%%%%%%%55

%%%%%%%%%%%%%%%%%%%%%%%%%%%%%%%%%%%%%%%%%%%%%%%%%%%%%%%%%%%%%%%%%%%%%%%%%%%%%%%%%

\subsection{Research Gaps and Study Positioning}

Although prior studies have advanced the understanding of adversarial attacks and defenses for ML-based NIDS, several limitations remain. First, many studies focus on a limited set of attack methods, models, or defense mechanisms. This makes it difficult to compare the relative effectiveness of different attacks and defenses under a common evaluation setting. For example, some studies focus primarily on FGSM or PGD, while others consider only one or two model families. As a result, the consistency of attack success across heterogeneous detection models remains insufficiently explored.

Second, some prior evaluations rely on older or single-source datasets, which may limit the relevance of the findings to more recent and diverse network environments~\cite{debicha2021adversarial,schneider2021evaluating,rigaki2017adversarial,debicha12detect}. Dataset choice is particularly important in NIDS because traffic distributions, feature representations, and attack behaviors vary across network environments. In this work, we use NF-UQ-NIDS~\cite{sarhan2021netflow}, a unified NetFlow-based benchmark constructed from multiple NIDS datasets. This provides broader traffic and attack coverage than a single-source dataset while maintaining a consistent NetFlow feature representation.

Third, robustness evaluation is often reported using isolated metrics, such as clean-data accuracy or attack success rate. While these metrics are informative, they do not always capture the trade-off between detection performance and resistance to adversarial evasion. A model may perform well on clean samples but remain vulnerable to adversarial examples, or it may reduce attack success while suffering degraded classification performance. This motivates the need for a compact comparative metric that considers both detection performance and evasion resistance.

Fourth, existing defense evaluations often focus on one defense mechanism at a time. However, adversarial training, ensemble learning, and adversarial detection may behave differently depending on the attack type and model architecture. Therefore, evaluating representative defense strategies under the same transfer-based setting is necessary to understand their relative strengths and limitations.

This study addresses these gaps by conducting a comparative empirical evaluation of eight adversarial evasion attacks, fifteen detection models, and three representative defense strategies using NF-UQ-NIDS. The evaluated attacks include both white-box generation and black-box transfer settings, allowing us to examine how adversarial samples generated against a surrogate model affect other ML-based NIDS models. We also introduce the Robustness Index (RI), a compact comparative metric that combines balanced accuracy and macro-F1 score computed on the robustness evaluation set with attack success rate. Finally, we present AR-NIDS, a two-stage framework that uses an adversarially trained ensemble to distinguish normal, attack, and adversarial samples, followed by an adversarial attack classifier to identify the attack type. Under the evaluated transfer-based setting, this framework provides an empirical comparison of how combined adversarial training and ensemble-based modeling can improve robustness against transferable adversarial examples.
%%%%%%%%%%%%%%%%%%%%%%%%%%%%%%%%%%%%%%%%%%%%%%%%%%%%%%%%%%%%%%%%%%%%
%%%%%%%%%%%%%%%%%%%%%%%%%%%%%%%%%%%%%%%%%%%%%%%%%%%%%%%%%%%%%%%%%%%

\section{Framework and Index for ML-based NIDS Robustness}
\label{Framework and Index for ML-based NIDS Robustness}

This section introduces two components used in the robustness evaluation. First, we define the Robustness Index (RI), a compact comparative metric for comparing ML-based NIDS models when adversarial samples are included in the evaluation. Second, we describe the Adversarial-Robust NIDS (AR-NIDS) framework, which combines adversarial training and ensemble-based learning with an adversarial attack classifier.

\subsection{Robustness Index}
\label{Robustness Index}

Robustness evaluation in ML-based NIDS should consider both classification performance and resistance to adversarial evasion. Reporting a single metric such as accuracy can be misleading, especially in imbalanced intrusion detection datasets. Similarly, reporting attack success rate (ASR) alone does not show whether a model maintains reliable classification performance when adversarial samples are included in the evaluation. Therefore, we introduce the Robustness Index (RI) as a compact comparative metric that summarizes the trade-off between performance on the robustness evaluation set and resistance to adversarial evasion.

The RI combines three indicators: balanced accuracy computed on the robustness evaluation set (\(\text{BA}_{R}\)), macro-F1 score computed on the robustness evaluation set (\(\text{F1}_{R}\)), and attack success rate (ASR). In this study, the robustness evaluation set refers to the test setting in which adversarial samples are evaluated together with clean normal and attack samples. For the baseline binary evaluation, adversarial samples are generated from attack records and evaluated as malicious samples. For the AR-NIDS defense evaluation, adversarial samples are included as a separate \textit{Adversarial} class. Thus, \(\text{BA}_{R}\) and \(\text{F1}_{R}\) measure classification performance when adversarial samples are present, while ASR measures how often adversarial samples evade detection.

Balanced accuracy is included because NIDS datasets are often imbalanced, with normal traffic usually outnumbering attack traffic. It gives equal importance to the evaluated classes by averaging class-level recall. For \(C\) evaluated classes, it is defined as:

\begin{equation}
\text{BA}_{R} = \frac{1}{C}\sum_{c=1}^{C}
\frac{\text{TP}_{c}}{\text{TP}_{c}+\text{FN}_{c}}
\end{equation}

where \(C\) is the number of classes in the evaluation setting. In the binary baseline setting, \(C=2\) for \textit{Normal} and \textit{Attack}; in the AR-NIDS defense setting, \(C=3\) for \textit{Normal}, \textit{Attack}, and \textit{Adversarial}.

Macro-F1 score is included because it provides a balanced view of precision and recall across classes and reduces the dominance of the majority class. In the robustness evaluation, it reflects the model's ability to maintain classification quality when adversarial samples are included:

\begin{equation}
\text{F1}_{R} = \frac{1}{C}\sum_{c=1}^{C}
2 \times \frac{\text{Precision}_{c}\times \text{Recall}_{c}}
{\text{Precision}_{c}+\text{Recall}_{c}}
\end{equation}

ASR measures the proportion of adversarial samples that successfully evade detection. In this study, an adversarial sample is considered successful when it is classified as normal. Therefore, ASR is negatively associated with robustness:

\begin{equation}
\text{ASR} = 
\frac{\text{Number of successful adversarial samples}}
{\text{Total number of adversarial samples}}
\end{equation}

The Robustness Index is then defined as:

\begin{equation}
\text{RI} = 
\frac{(\text{BA}_{R} + \text{F1}_{R} - \text{ASR}) + 1}{3}
\end{equation}

The RI uses an equal-weight linear combination of the three components. The positive terms, \(\text{BA}_{R}\) and \(\text{F1}_{R}\), increase the index when classification performance on the robustness evaluation set improves, while the negative term, ASR, decreases the index when adversarial evasion succeeds. The addition of 1 shifts the minimum possible value of the numerator from \(-1\) to 0, and division by 3 normalizes the index to the interval \([0,1]\), assuming all component metrics are bounded between 0 and 1.

In the ideal case, \(\text{BA}_{R}=1\), \(\text{F1}_{R}=1\), and \(\text{ASR}=0\). Substituting these values gives:

\[
\text{RI} = \frac{(1 + 1 - 0) + 1}{3} = 1
\]

This represents the best possible score under the proposed index. In the worst case, \(\text{BA}_{R}=0\), \(\text{F1}_{R}=0\), and \(\text{ASR}=1\). Substituting these values gives:

\[
\text{RI} = \frac{(0 + 0 - 1) + 1}{3} = 0
\]

This represents the lowest possible score under the proposed index.

The purpose of RI is not to replace the individual metrics, but to support model comparison when multiple robustness-related metrics must be considered together. Two models may have similar classification performance but different ASR values, or similar ASR values but different performance when adversarial samples are included. RI provides a compact way to compare these trade-offs while retaining interpretability. For this reason, the experimental results report RI together with its underlying performance metrics rather than using it as a standalone proof of robustness.

\subsection{Adversarial-Robust Network Intrusion Detection System (AR-NIDS)}
\label{Adversarial-Robust Network Intrusion Detection System (AR-NIDS)}

\begin{figure*}[!t]
\centering
\fbox{\includegraphics[width=0.60\textwidth]{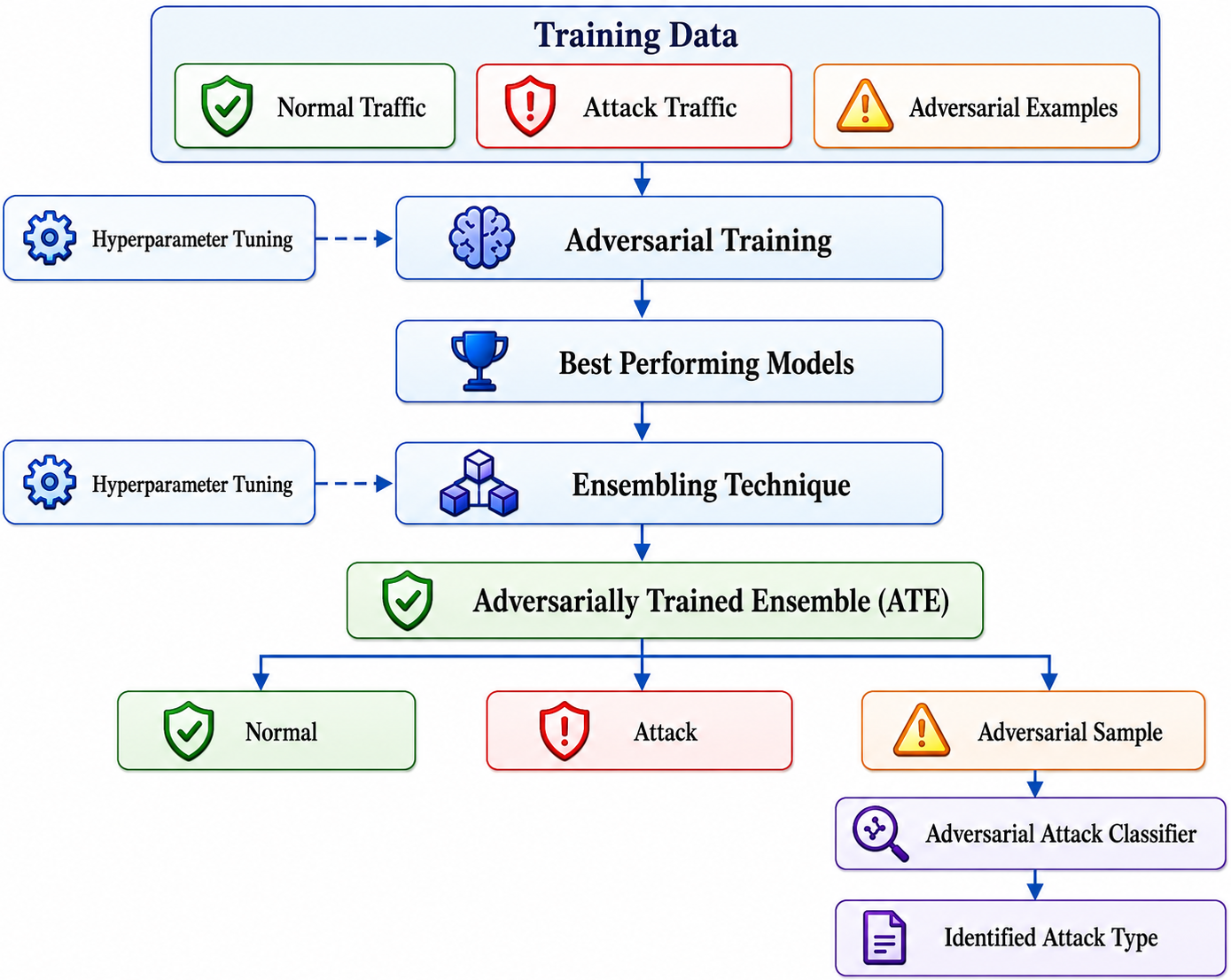}}
\caption{Adversarial-Robust NIDS framework.}
\label{Adversarial-Robust Network Intrusion Detection System}
\end{figure*}

This section presents the Adversarial-Robust NIDS (AR-NIDS) framework used in this study. AR-NIDS is designed as a two-stage framework for detecting adversarially perturbed samples and identifying the attack method associated with them. Figure~\ref{Adversarial-Robust Network Intrusion Detection System} illustrates the main components of the framework.

The first stage is the Adversarially Trained Ensemble (ATE) detector. The ATE detector combines adversarial training with ensemble-based learning. Adversarial training exposes the detector to clean and adversarial samples during training, allowing it to learn patterns associated with adversarial perturbations. Ensemble-based learning reduces dependence on a single decision boundary and can improve stability by aggregating decisions across multiple learners.

In this study, the ATE detector is implemented using an LGBM-based ensemble and is trained as a multiclass classifier with three labels: normal, attack, and adversarial. The normal class represents benign samples, the attack class represents non-adversarial malicious samples, and the adversarial class represents malicious samples modified by adversarial evasion techniques. This design allows the detector to distinguish between ordinary attacks and adversarially perturbed attacks rather than treating all malicious samples as a single category.

The second stage is the adversarial attack classifier. When a sample is identified as adversarial by the first stage, it is passed to this classifier to identify the adversarial attack type. The classifier is implemented using LGBM and is trained to distinguish among the evaluated attack methods, including FGSM, PGD, JSMA, DeepFool, CW2, CW\(\infty\), ZOO, and HSJ. This stage supports attack characterization by indicating which adversarial method most likely generated the detected sample.

The AR-NIDS framework should be interpreted within the evaluation scope of this study. It is designed and evaluated under a transfer-based adversarial setting, where adversarial samples are generated against a surrogate model and assessed against the evaluated detection models. Therefore, AR-NIDS demonstrates empirical robustness under the evaluated setting rather than a universal robustness guarantee. Its purpose is to examine whether combining adversarial training with ensemble-based modeling can improve robustness against transferable adversarial samples while also supporting adversarial attack-type classification.
%%%%%%%%%%%%%%%%%%%%%%%%%%%%%%%%%%%%%%%%%%%%%%%%%%%%%%%%%%%%%%%%%%%%%%%

\section{Experimental Setup}\label{experimental setup}

This section describes the experimental setting used to evaluate ML-based NIDS robustness against transferable adversarial examples. It presents the adversary model, dataset, preprocessing steps, adversarial example generation process, detection models, and evaluation metrics.

\subsection{Adversary Model}\label{adversary model}

We consider an evasion adversary whose goal is to perturb malicious samples so that they are misclassified as normal by the NIDS. The adversarial examples are generated from malicious records and are used to evaluate whether perturbations crafted against one model can transfer to other detection models. In this study, adversarial examples are generated in the normalized NetFlow feature space, which supports controlled comparison of model sensitivity and attack transferability under a common evaluation setting. The evaluation therefore represents feature-level robustness analysis rather than packet-level attack realizability.

The evaluation follows a transfer-based setting. Adversarial examples are first generated against a Feed-Forward Deep Neural Network (FF-DNN), which acts as the surrogate model. This corresponds to a white-box generation step because the attacks are crafted using the surrogate model. The generated adversarial examples are then evaluated against the remaining trained ML-based NIDS models. This corresponds to a black-box transfer evaluation because those target models are not used during adversarial example generation.

The label setting differs according to the evaluation stage. For the baseline robustness evaluation, the original NIDS task is formulated as a binary classification problem with two labels: \textit{Normal} and \textit{Attack}. In this setting, adversarial examples generated from attack records are evaluated as attack samples, because the goal is to test whether perturbed malicious samples evade detection by being classified as normal. For the defense evaluation and the proposed AR-NIDS framework, the task is extended to a three-class setting with labels \textit{Normal}, \textit{Attack}, and \textit{Adversarial}. In this setting, adversarially perturbed attack samples are assigned to the \textit{Adversarial} class so that the model can explicitly learn to distinguish ordinary attacks from adversarially perturbed attacks.

The reported results should therefore be interpreted as empirical robustness under the evaluated transfer-based setting and attack set.

\subsection{Dataset}\label{dataset}

Numerous publicly accessible datasets have been used to develop and assess ML-based NIDS. However, differences in feature sets and traffic-generation environments can make direct comparison across datasets difficult. To address this issue, Sarhan et al.~\cite{sarhan2021netflow} introduced NetFlow-based NIDS datasets with consistent and practical flow-level features. In this study, we use NF-UQ-NIDS~\cite{sarhan2021netflow}, a unified NetFlow-based benchmark constructed from multiple NIDS datasets. This provides broader traffic and attack coverage than a single-source dataset while maintaining a consistent feature representation.

The full NF-UQ-NIDS dataset contains 11,994,893 records, including 9,208,048 benign records (76.77\%) and 2,786,845 attack records (23.23\%). It includes twenty distinct attack categories. In our experiments, we use a subset of 6,338,509 records, consisting of 81.8\% normal records and 18.2\% attack records. Table~\ref{Statistics of Used NF-UQ-NIDS Dataset} summarizes the distribution of traffic types in the used subset.

For the baseline NIDS evaluation, all original attack categories are unified under a single \textit{Attack} label, resulting in a binary classification task with two classes: \textit{Normal} and \textit{Attack}. The dataset is stratified into training (80\%) and testing (20\%) subsets based on class labels. For the AR-NIDS defense evaluation, adversarial samples generated from attack records are added as a third class labeled \textit{Adversarial}. Table~\ref{Samples Distribution in Training and Testing Data} reports the final sample distribution used in the evaluation.

\begin{table}[!t]
\scriptsize
\centering
\setlength{\tabcolsep}{4pt}
\renewcommand{\arraystretch}{0.88}
\begin{tabular}{|l|c|c|}
\hline
\textbf{Flow Type} & \textbf{Number of Samples} & \textbf{Ratio \%} \\ \hline 
Benign            & 5,184,112 & 81.7727 \\ \hline
DDoS              & 151,866   & 2.3942  \\ \hline
password          & 151,866   & 2.3942  \\ \hline
injection         & 151,866   & 2.3942  \\ \hline
xss               & 151,866   & 2.3942  \\ \hline
DoS               & 151,866   & 2.3942  \\ \hline
Reconnaissance    & 151,866   & 2.3942  \\ \hline
scanning          & 151,866   & 2.3942  \\ \hline
Brute Force       & 27,721    & 0.4372  \\ \hline
Infiltration      & 26,131    & 0.4121  \\ \hline
Bot               & 16,829    & 0.2655  \\ \hline
Exploits          & 6,591     & 0.1039  \\ \hline
Fuzzers           & 4,462     & 0.0704  \\ \hline
Backdoor          & 3,856     & 0.0608  \\ \hline
Generic           & 2,268     & 0.0358  \\ \hline
mitm              & 1,723     & 0.0272  \\ \hline
ransomware        & 784       & 0.0124  \\ \hline
Theft             & 533       & 0.0084  \\ \hline
Shellcode         & 312       & 0.0049  \\ \hline
Analysis          & 90        & 0.0014  \\ \hline
Worms             & 35        & 0.0006  \\ \hline\hline  
\textbf{Total}    & 6,338,509 & 100     \\ \hline 
\end{tabular}
\renewcommand{\arraystretch}{1}
\caption{Statistics of the used NF-UQ-NIDS subset.}
\label{Statistics of Used NF-UQ-NIDS Dataset}
\end{table}

\begin{table}[!t]
\scriptsize
\centering
\renewcommand{\arraystretch}{0.92}
\begin{tabular}{lcc}
\toprule
\textbf{Samples} & \textbf{Train} & \textbf{Test} \\
\midrule
Normal & 4,147,289 & 1,036,823 \\
Attack & 923,518 & 230,879 \\
Adversarial & 820,903 & 205,224 \\
\bottomrule
\end{tabular}
\renewcommand{\arraystretch}{1}
\caption{Sample distribution in the training and testing sets.}
\label{Samples Distribution in Training and Testing Data}
\end{table}

\subsection{Dataset Preprocessing}\label{Dataset Pre-processing}

The features were transformed using Min-Max normalization, rescaling each feature to the range \([0,1]\). This preprocessing step places all features on a common scale and prevents features with larger numeric ranges from dominating model training or adversarial perturbation generation. Min-Max scaling is defined as:

\begin{equation}
X_{norm}=\frac{X-X_{min}}{X_{max}-X_{min}}
\end{equation}

where \(X\) is the original feature value, and \(X_{min}\) and \(X_{max}\) are the minimum and maximum values of the corresponding feature.

\subsection{Adversarial Attacks and Models}\label{Adversarial Attacks and Models}

The Adversarial Robustness Toolbox (ART) library~\cite{nicolae2018adversarial} was used to generate adversarial examples. Eight evasion attack techniques were evaluated: FGSM, PGD, HSJ, JSMA, DeepFool, CW2, ZOO, and CW\(\infty\). These attacks represent gradient-based, optimization-based, and query-based adversarial methods.

Adversarial examples were generated from malicious records using the FF-DNN model as the surrogate model. Across the eight evaluated attacks, the aggregated adversarial set contains 820,903 training samples and 205,224 testing samples, as reported in Table~\ref{Samples Distribution in Training and Testing Data}. These adversarial samples are used to evaluate performance degradation, attack success, and the effectiveness of defense strategies.

A consistent ART configuration was used across the evaluated attacks to provide a comparable benchmark. This design supports controlled comparison across attack methods and models. Therefore, the reported attack success rates should be interpreted under this common experimental configuration.

The FF-DNN was implemented using Keras with TensorFlow backend. Bayesian Optimization was used to tune the FF-DNN architecture and training configuration. In addition to the FF-DNN, fourteen conventional ML algorithms were evaluated, resulting in fifteen detection models in total. The conventional models include Random Forest (RF), Decision Tree (DT), Extra Trees (ET), Logistic Regression (LR), eXtreme Gradient Boosting (XGBoost), CatBoost, Gradient Boosting Decision Tree (GBDT), Light Gradient Boosting Machine (LGBM), Adaptive Boosting (AdaBoost), Multilayer Perceptron (MLP), Multinomial Naive Bayes (NB), Linear Discriminant Analysis (LDA), Quadratic Discriminant Analysis (QDA), and Stochastic Gradient Descent (SGD). These models were selected because they represent diverse model families commonly used in ML-based intrusion detection, including tree-based, ensemble-based, linear, probabilistic, and neural models.

For reproducibility, the model parameters used in the experiments are summarized in Table~\ref{Parameters of the ML-Based NIDS Models}. The implementation and supporting experimental files are publicly available on Zenodo.\footnote{\url{https://doi.org/10.5281/zenodo.20286264}}

\begin{figure*}[!t]
\centering
\fbox{\includegraphics[width=0.9\textwidth]{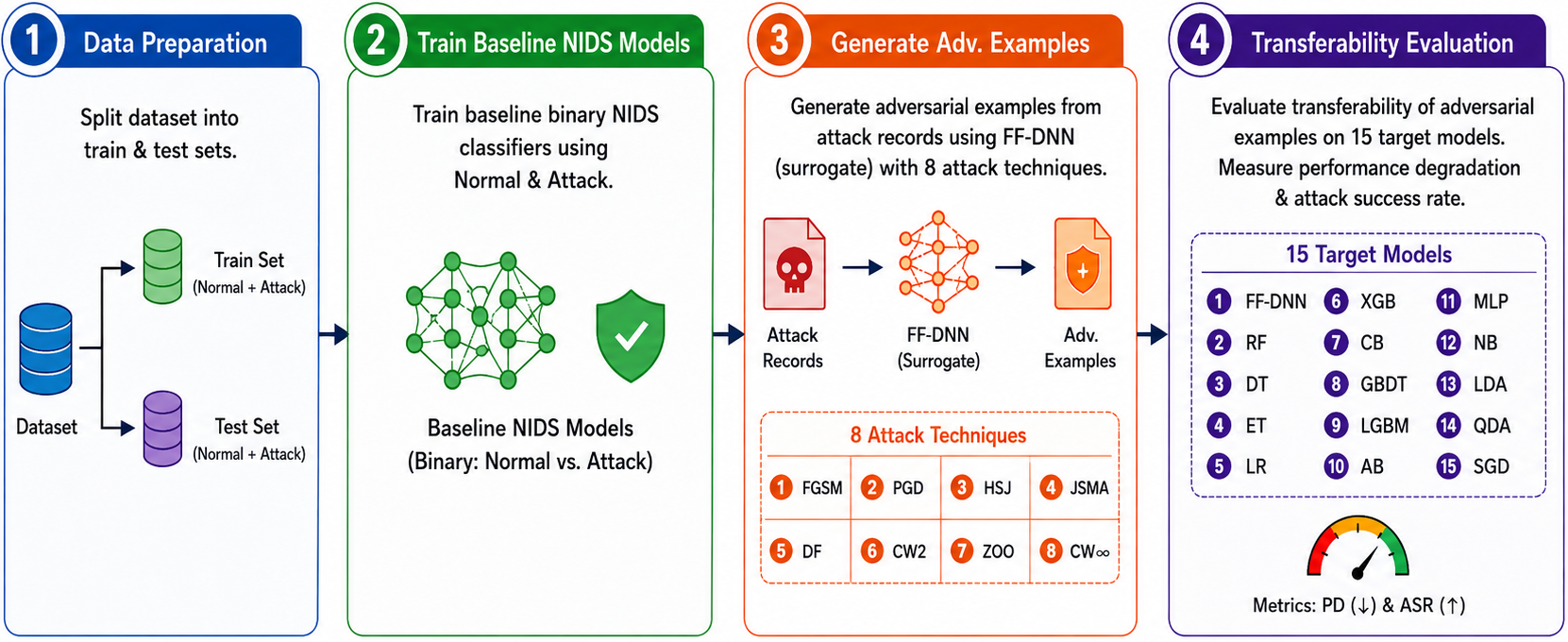}}
\caption{Transfer-based evaluation of ML-NIDS robustness to adversarial attacks.}
\label{Assessment of ML-based NIDS Robustness to Transferable Adversarial Attacks}
\end{figure*}

The transferability assessment workflow is summarized in Fig.~\ref{Assessment of ML-based NIDS Robustness to Transferable Adversarial Attacks}, including dataset splitting, baseline model training, adversarial example generation, and evaluation across target models.

\begin{table}[!t]
\small
\centering
\renewcommand{\arraystretch}{0.90}
\resizebox{\columnwidth}{!}{%
\begin{tabular}{|l|l|}
\hline
\textbf{Model} & \textbf{Parameters} \\ \hline
FF-DNN & \begin{tabular}[c]{@{}l@{}}Layers: [Dense(128, activation=relu), Dropout, Dense(64, activation=relu), Dropout,\\ Dense(2, activation=softmax)], Optimizer: Adam\end{tabular} \\ \hline
RF & n\_estimators=100, criterion='gini', max\_depth=None \\ \hline
DT & criterion='gini', splitter='best', max\_depth=None \\ \hline
ET & n\_estimators=100, criterion='gini', max\_depth=None \\ \hline
LR & penalty='l2', C=1.0, solver='lbfgs' \\ \hline
XGBoost & objective='binary:logistic', n\_estimators=100, max\_depth=3 \\ \hline
AdaBoost & base\_estimator=None, n\_estimators=50, learning\_rate=1.0 \\ \hline
NB & alpha=1.0, fit\_prior=True, class\_prior=None \\ \hline
LDA & solver='svd', shrinkage=None, priors=None \\ \hline
QDA & priors=None, reg\_param=0.0, store\_covariance=False \\ \hline
SGD & loss='hinge', penalty='l2', alpha=0.0001 \\ \hline
GBDT & n\_estimators=100, learning\_rate=0.1, max\_depth=3, min\_samples\_split=2 \\ \hline
CatBoost & iterations=100, depth=3, learning\_rate=0.03, boosting\_type='Ordered'\\ \hline
LGBM & n\_estimators=100, learning\_rate=0.1, max\_depth=-1, boosting\_type='gbdt' \\ \hline
MLP & hidden\_layer\_sizes=(100,), activation='relu', solver='adam', learning\_rate\_init=0.001 \\ \hline
\end{tabular}
}
\renewcommand{\arraystretch}{1}
\caption{Parameters of the ML-based NIDS models.}
\label{Parameters of the ML-Based NIDS Models}
\end{table}

\subsection{Models' Performance Evaluation Metrics}\label{Models Performance Evaluation Metrics}

Because NIDS datasets are often imbalanced, standard accuracy alone can be misleading. Therefore, model performance is evaluated using balanced accuracy, precision, recall, macro F1-score, attack success rate, and the Robustness Index described in Sections~\ref{Evaluation Metrics for ML-based NIDS} and~\ref{Robustness Index}.

Balanced accuracy is used to account for class imbalance by giving equal importance to attack and normal classes. Recall for the attack class is important because it measures the model's ability to detect malicious samples; a low attack recall indicates that attacks are being missed. Precision for the normal class is considered because adversarial evasion aims to make malicious or adversarial samples appear normal. Macro F1-score is used because it averages class-level F1-scores without weighting by class frequency, providing a balanced measure across classes.

For adversarial robustness, ASR measures the proportion of adversarial samples that successfully evade detection. RI is then used as a compact comparative metric that combines balanced accuracy and macro-F1 score computed on the robustness evaluation set with ASR. The metrics are reported together to avoid relying on a single measure and to make the trade-off between detection performance and adversarial evasion resistance explicit.
%%%%%%%%%%%%%%%%%%%%%%%%%%%%%%%%%%%%%%%%%%%%%%%%%%%%%%%%%%%%%%%%%%%%%%%

\section{Robustness Evaluation of ML-based NIDS to Adversarial Attacks}
\label{Robustness Evaluation of ML-based NIDS to Adversarial Attacks}

This section evaluates the robustness of baseline ML-based NIDS models under the transfer-based adversarial setting described in Section~\ref{experimental setup}. The evaluation considers two aspects: model performance on clean test data and model performance when adversarial examples generated against the FF-DNN surrogate model are introduced. We further analyze robustness using the Robustness Index (RI) and examine the consistency of attack success across the evaluated models.

\subsection{Model Performance on Clean Test Data}
\label{Models Performance Over Clean Test Data}

Table~\ref{Baseline Models Performance over Clean vs Adversarial Data} reports the performance of the evaluated models on clean test data and on the robustness evaluation set. This subsection first discusses the clean-data results to establish the baseline performance of each model before adversarial perturbations are introduced.

On clean test data, ensemble and tree-based models achieve the strongest overall performance. CatBoost, DT, ET, RF, and XGBoost obtain balanced accuracy values of 0.98, indicating strong ability to distinguish normal and attack samples under the original binary NIDS setting. MLP, FF-DNN, and GBDT also perform strongly, with macro F1-scores around 0.95 or higher. In contrast, linear and probabilistic models show lower performance. SGD, LR, LDA, NB, and QDA obtain weaker balanced accuracy and macro F1-scores, with QDA achieving the lowest macro F1-score of 0.46.

These results indicate that ensemble and tree-based models are better suited to the clean NF-UQ-NIDS feature space than simpler linear or distribution-based classifiers. This may be explained by their ability to model nonlinear relationships and feature interactions in network flow data. However, strong clean-data performance does not necessarily imply adversarial robustness, as shown in the following subsection.

\begin{table*}[!t]
\scriptsize
\centering
\setlength{\tabcolsep}{3pt}
\renewcommand{\arraystretch}{0.92}
\begin{tabular}{|l|ccc|ccc|ccc|ccc|}
\hline
\multirow{2}{*}{\textbf{Model}} & \multicolumn{3}{c|}{\textbf{Balanced Accuracy}} & \multicolumn{3}{c|}{\textbf{Precision (Normal)}} & \multicolumn{3}{c|}{\textbf{Recall (Attack)}} & \multicolumn{3}{c|}{\textbf{Macro F1-score}} \\ \cline{2-13} 
& \multicolumn{1}{c|}{\textbf{clean}} & \multicolumn{1}{c|}{\textbf{adv.}} & \textbf{Drop} 
& \multicolumn{1}{c|}{\textbf{clean}} & \multicolumn{1}{c|}{\textbf{adv.}} & \textbf{Drop} 
& \multicolumn{1}{c|}{\textbf{clean}} & \multicolumn{1}{c|}{\textbf{adv.}} & \textbf{Drop} 
& \multicolumn{1}{c|}{\textbf{clean}} & \multicolumn{1}{c|}{\textbf{adv.}} & \textbf{Drop} \\ \hline
\textbf{QDA}      & \multicolumn{1}{c|}{0.66} & \multicolumn{1}{c|}{0.48} & 0.18 & \multicolumn{1}{c|}{0.98} & \multicolumn{1}{c|}{0.68} & 0.30 & \multicolumn{1}{c|}{0.97} & \multicolumn{1}{c|}{0.61} & 0.36 & \multicolumn{1}{c|}{0.46} & \multicolumn{1}{c|}{0.43} & 0.03 \\ \hline
\textbf{FFDNN}    & \multicolumn{1}{c|}{0.95} & \multicolumn{1}{c|}{0.78} & 0.17 & \multicolumn{1}{c|}{0.98} & \multicolumn{1}{c|}{0.84} & 0.14 & \multicolumn{1}{c|}{0.92} & \multicolumn{1}{c|}{0.57} & 0.35 & \multicolumn{1}{c|}{0.95} & \multicolumn{1}{c|}{0.81} & 0.14 \\ \hline
\textbf{RF}       & \multicolumn{1}{c|}{0.98} & \multicolumn{1}{c|}{0.82} & 0.16 & \multicolumn{1}{c|}{0.99} & \multicolumn{1}{c|}{0.87} & 0.12 & \multicolumn{1}{c|}{0.97} & \multicolumn{1}{c|}{0.65} & 0.32 & \multicolumn{1}{c|}{0.99} & \multicolumn{1}{c|}{0.86} & 0.13 \\ \hline
\textbf{CatBoost} & \multicolumn{1}{c|}{0.98} & \multicolumn{1}{c|}{0.84} & 0.14 & \multicolumn{1}{c|}{0.99} & \multicolumn{1}{c|}{0.88} & 0.11 & \multicolumn{1}{c|}{0.97} & \multicolumn{1}{c|}{0.69} & 0.28 & \multicolumn{1}{c|}{0.98} & \multicolumn{1}{c|}{0.87} & 0.11 \\ \hline
\textbf{DT}       & \multicolumn{1}{c|}{0.98} & \multicolumn{1}{c|}{0.84} & 0.14 & \multicolumn{1}{c|}{0.99} & \multicolumn{1}{c|}{0.88} & 0.11 & \multicolumn{1}{c|}{0.97} & \multicolumn{1}{c|}{0.69} & 0.28 & \multicolumn{1}{c|}{0.98} & \multicolumn{1}{c|}{0.87} & 0.11 \\ \hline
\textbf{ET}       & \multicolumn{1}{c|}{0.98} & \multicolumn{1}{c|}{0.84} & 0.14 & \multicolumn{1}{c|}{0.99} & \multicolumn{1}{c|}{0.88} & 0.11 & \multicolumn{1}{c|}{0.97} & \multicolumn{1}{c|}{0.68} & 0.29 & \multicolumn{1}{c|}{0.98} & \multicolumn{1}{c|}{0.87} & 0.11 \\ \hline
\textbf{XGB}      & \multicolumn{1}{c|}{0.98} & \multicolumn{1}{c|}{0.86} & 0.12 & \multicolumn{1}{c|}{0.99} & \multicolumn{1}{c|}{0.90} & 0.09 & \multicolumn{1}{c|}{0.97} & \multicolumn{1}{c|}{0.74} & 0.23 & \multicolumn{1}{c|}{0.98} & \multicolumn{1}{c|}{0.89} & 0.09 \\ \hline
\textbf{AdaBoost} & \multicolumn{1}{c|}{0.93} & \multicolumn{1}{c|}{0.84} & 0.09 & \multicolumn{1}{c|}{0.98} & \multicolumn{1}{c|}{0.90} & 0.08 & \multicolumn{1}{c|}{0.93} & \multicolumn{1}{c|}{0.75} & 0.18 & \multicolumn{1}{c|}{0.89} & \multicolumn{1}{c|}{0.85} & 0.04 \\ \hline
\textbf{GBDT}     & \multicolumn{1}{c|}{0.96} & \multicolumn{1}{c|}{0.87} & 0.09 & \multicolumn{1}{c|}{0.99} & \multicolumn{1}{c|}{0.91} & 0.08 & \multicolumn{1}{c|}{0.94} & \multicolumn{1}{c|}{0.76} & 0.18 & \multicolumn{1}{c|}{0.95} & \multicolumn{1}{c|}{0.89} & 0.06 \\ \hline
\textbf{LR}       & \multicolumn{1}{c|}{0.81} & \multicolumn{1}{c|}{0.75} & 0.06 & \multicolumn{1}{c|}{0.94} & \multicolumn{1}{c|}{0.86} & 0.08 & \multicolumn{1}{c|}{0.78} & \multicolumn{1}{c|}{0.67} & 0.11 & \multicolumn{1}{c|}{0.75} & \multicolumn{1}{c|}{0.75} & 0.00 \\ \hline
\textbf{MLP}      & \multicolumn{1}{c|}{0.95} & \multicolumn{1}{c|}{0.89} & 0.06 & \multicolumn{1}{c|}{0.98} & \multicolumn{1}{c|}{0.92} & 0.06 & \multicolumn{1}{c|}{0.91} & \multicolumn{1}{c|}{0.80} & 0.11 & \multicolumn{1}{c|}{0.96} & \multicolumn{1}{c|}{0.91} & 0.05 \\ \hline
\textbf{LGBM}     & \multicolumn{1}{c|}{0.96} & \multicolumn{1}{c|}{0.91} & 0.05 & \multicolumn{1}{c|}{1.00} & \multicolumn{1}{c|}{0.95} & 0.05 & \multicolumn{1}{c|}{0.98} & \multicolumn{1}{c|}{0.87} & 0.11 & \multicolumn{1}{c|}{0.92} & \multicolumn{1}{c|}{0.90} & 0.02 \\ \hline
\textbf{SGD}      & \multicolumn{1}{c|}{0.83} & \multicolumn{1}{c|}{0.78} & 0.05 & \multicolumn{1}{c|}{0.95} & \multicolumn{1}{c|}{0.88} & 0.07 & \multicolumn{1}{c|}{0.82} & \multicolumn{1}{c|}{0.73} & 0.09 & \multicolumn{1}{c|}{0.76} & \multicolumn{1}{c|}{0.77} & -0.01 \\ \hline
\textbf{LDA}      & \multicolumn{1}{c|}{0.69} & \multicolumn{1}{c|}{0.66} & 0.03 & \multicolumn{1}{c|}{0.88} & \multicolumn{1}{c|}{0.78} & 0.10 & \multicolumn{1}{c|}{0.43} & \multicolumn{1}{c|}{0.36} & 0.07 & \multicolumn{1}{c|}{0.72} & \multicolumn{1}{c|}{0.68} & 0.04 \\ \hline
\textbf{NB}       & \multicolumn{1}{c|}{0.70} & \multicolumn{1}{c|}{0.67} & 0.03 & \multicolumn{1}{c|}{0.89} & \multicolumn{1}{c|}{0.80} & 0.09 & \multicolumn{1}{c|}{0.54} & \multicolumn{1}{c|}{0.48} & 0.06 & \multicolumn{1}{c|}{0.69} & \multicolumn{1}{c|}{0.68} & 0.01 \\ \hline\hline
\textbf{Average}  & \multicolumn{1}{c|}{0.89} & \multicolumn{1}{c|}{0.79} & 0.10 & \multicolumn{1}{c|}{0.97} & \multicolumn{1}{c|}{0.86} & 0.11 & \multicolumn{1}{c|}{0.87} & \multicolumn{1}{c|}{0.67} & 0.20 & \multicolumn{1}{c|}{0.86} & \multicolumn{1}{c|}{0.80} & 0.06 \\ \hline
\end{tabular}
\renewcommand{\arraystretch}{1}
\caption{Baseline model performance on clean test data and on the robustness evaluation set.}
\label{Baseline Models Performance over Clean vs Adversarial Data}
\end{table*}

\subsection{Performance Degradation under Adversarial Samples}
\label{Models Performance: Clean vs. Adversarial}

This subsection evaluates how adversarial examples generated against the FF-DNN surrogate affect the baseline detection models. In the baseline binary evaluation, adversarial examples are generated from attack records and evaluated as malicious samples. Therefore, a successful evasion occurs when a perturbed attack sample is classified as normal.

As shown in Table~\ref{Baseline Models Performance over Clean vs Adversarial Data}, adversarial samples reduce model performance across most metrics. The average balanced accuracy decreases from 0.89 on clean data to 0.79 on the robustness evaluation set. The largest balanced-accuracy drops are observed for QDA, FF-DNN, and RF, with reductions between 0.16 and 0.18. CatBoost, DT, ET, and XGB also show noticeable reductions between 0.12 and 0.14. In contrast, LR, MLP, LGBM, SGD, LDA, and NB show smaller drops between 0.03 and 0.06.

The most important degradation appears in attack recall. Average attack recall decreases from 0.87 on clean data to 0.67 on the robustness evaluation set, corresponding to a drop of 0.20. This indicates that adversarial perturbations reduce the ability of the models to detect malicious samples. QDA, FF-DNN, and RF show the largest recall reductions, ranging from 0.32 to 0.36. ET, CatBoost, DT, XGB, AdaBoost, and GBDT experience moderate recall reductions, while LR, MLP, LGBM, SGD, LDA, and NB show smaller recall drops.

The precision of the normal class also decreases, from an average of 0.97 on clean data to 0.86 on the robustness evaluation set. This suggests that the models become less reliable when distinguishing normal samples from adversarially perturbed malicious samples. QDA shows the largest reduction in normal precision, while LGBM and MLP maintain comparatively stronger performance. Overall, these results show that high clean-data performance is not sufficient to infer robustness against transferable adversarial examples.

\subsection{Model Robustness Based on RI}
\label{Models Adversarial Robustness}

\begin{figure*}[!t]
\centering
\fbox{\includegraphics[width=0.9\textwidth]{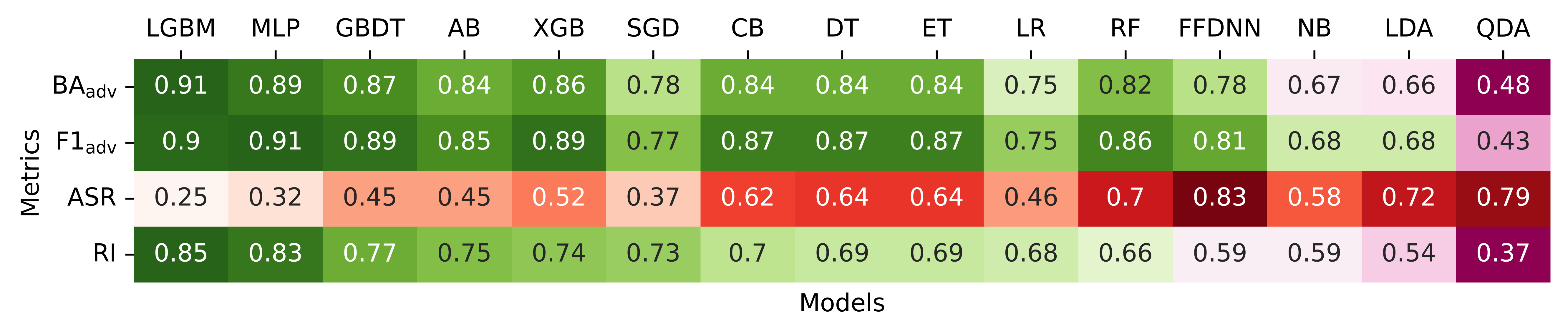}}
\caption{Baseline model Robustness Index values.}
\label{Baseline Models Robustness Index}
\end{figure*}

We use the Robustness Index (RI) introduced in Section~\ref{Robustness Index} to compare baseline models on the robustness evaluation set. RI combines balanced accuracy and macro-F1 score computed when adversarial samples are included with attack success rate into a compact comparative score. Higher RI values indicate a better trade-off between classification performance on the robustness evaluation set and resistance to evasion.

Figure~\ref{Baseline Models Robustness Index} shows that LGBM, MLP, and GBDT obtain the strongest RI values among the baseline models. LGBM achieves the highest RI, supported by high balanced accuracy and macro-F1 score on the robustness evaluation set, together with the lowest ASR among the baseline models. MLP and GBDT also maintain relatively strong RI values, indicating better robustness under the evaluated transfer setting.

CatBoost, DT, ET, RF, XGB, AdaBoost, LR, and SGD fall into a mid-range group. These models maintain reasonable performance on the robustness evaluation set but show higher attack success or larger performance degradation than the strongest models. This indicates that strong clean-data performance alone is insufficient for robust model selection.

NB, LDA, and QDA obtain the weakest RI values. Their lower RI scores reflect weaker classification performance when adversarial samples are included and, in some cases, higher vulnerability to evasion. These results suggest that models with simpler distributional or linear assumptions are less reliable under the evaluated adversarial setting.

\begin{figure*}[!t]
\centering
\begin{minipage}{.5\textwidth}
\centering
\includegraphics[width=0.9\textwidth]{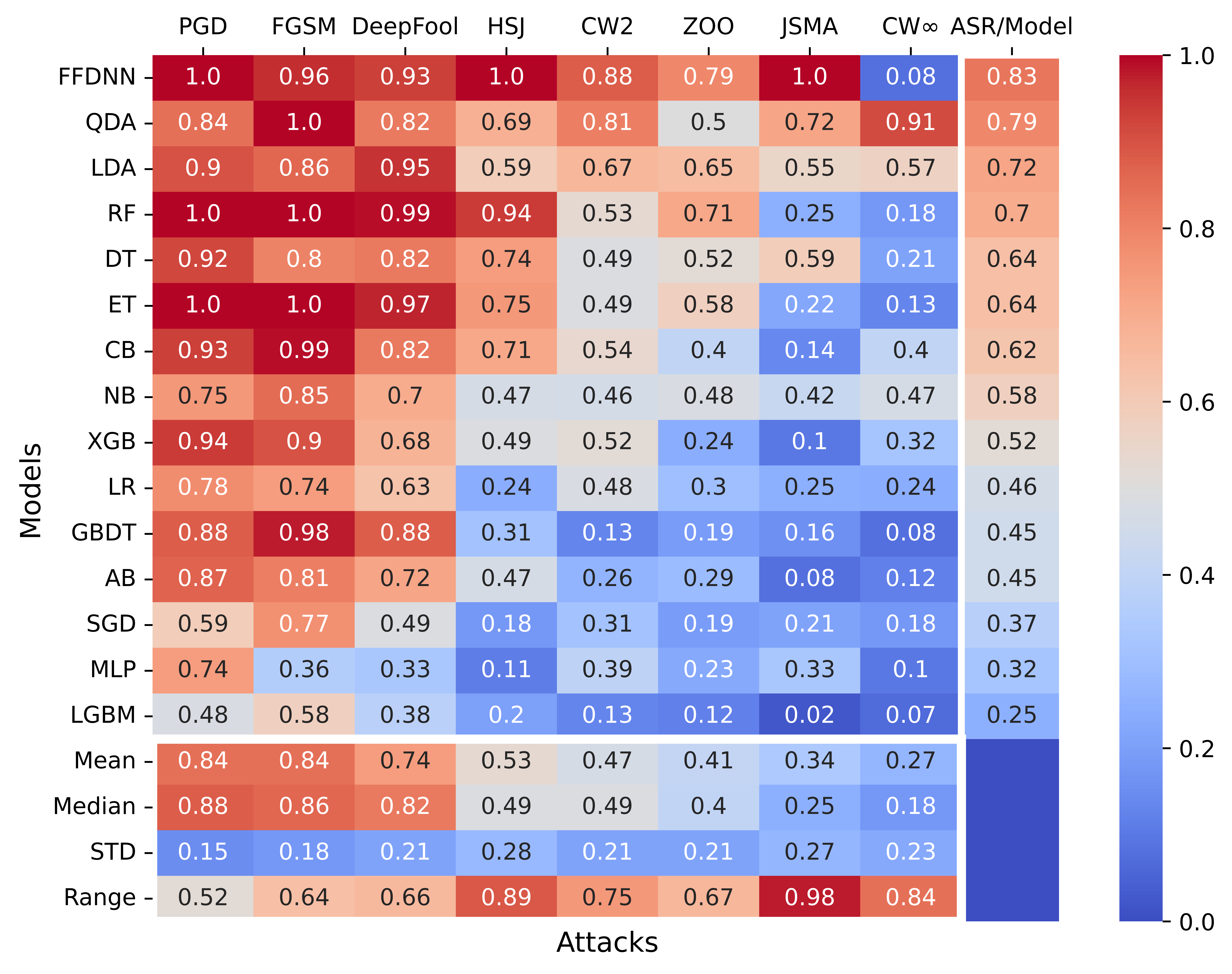}
\caption{Attack success rates against baseline models.}
\label{Attack Success Rates Against Baseline Models}
\end{minipage}%
\begin{minipage}{.5\textwidth}
\centering
\includegraphics[width=0.9\textwidth]{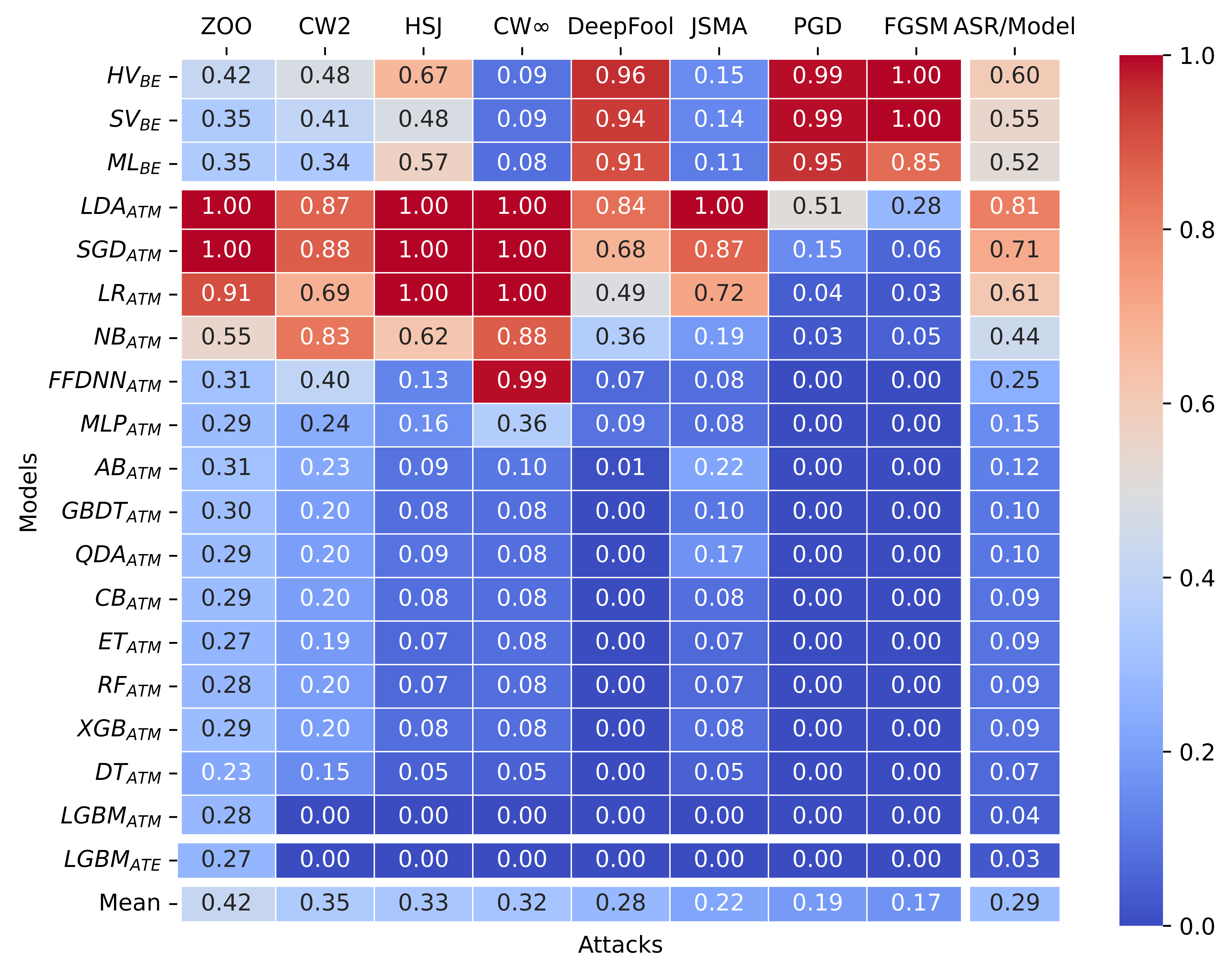}
\caption{Attack success rates against defense models.}
\label{Attack Success Rates Against Defense Models}
\end{minipage}
\end{figure*}

\subsection{Attack Transferability and Success Consistency}
\label{Adversarial Attacks Performance}

This subsection analyzes the transferability of the evaluated attacks across baseline models. We use the mean, median, standard deviation, and range of attack success rates to summarize both attack effectiveness and consistency. The mean and median indicate the overall level of attack success, while the standard deviation and range indicate how consistently an attack transfers across different models.

As shown in Figure~\ref{Attack Success Rates Against Baseline Models}, PGD, FGSM, and DeepFool are the most effective transferable attacks. Their median success rates range from 0.82 to 0.88, indicating that they remain effective across several baseline models. This suggests that the perturbations generated by these attacks against the FF-DNN surrogate can exploit vulnerabilities shared by multiple target models.

HSJ and CW2 show moderate transferability, with average success rates of 0.53 and 0.47, respectively. Their effectiveness varies more strongly across target models, indicating that transfer success depends on the target model structure. ZOO shows lower average transfer success, while JSMA and CW\(\infty\) show the lowest median success rates. These results indicate that not all adversarial methods transfer equally across ML-based NIDS models.

Overall, the attack-level analysis shows that adversarial risk in ML-based NIDS is attack-dependent and model-dependent. PGD, FGSM, and DeepFool represent the strongest transfer threats in this evaluation, while other attacks exhibit more variable success. This finding supports the need to evaluate multiple attack families rather than relying on a single adversarial method.

%%%%%%%%%%%%%%%%%%%%%%%%%%%%%%%%%%%%%%%%%%%%%%%%%%%%%%%%%%%%%%%%%%%
\section{Evaluation of Adversarial-Robust NIDS Framework}
\label{Evaluation of Adversarial-Robust NIDS}

This section evaluates the defense strategies considered in this study under the transfer-based adversarial setting. We compare three defense configurations: baseline ensemble learning (BE), adversarially trained models (ATM), and the proposed adversarially trained ensemble (ATE). The goal is to examine how each defense strategy affects classification performance on the robustness evaluation set, attack success rate, and the Robustness Index (RI).

\begin{figure*}[!t]
\centering
\fbox{\includegraphics[width=0.90\textwidth]{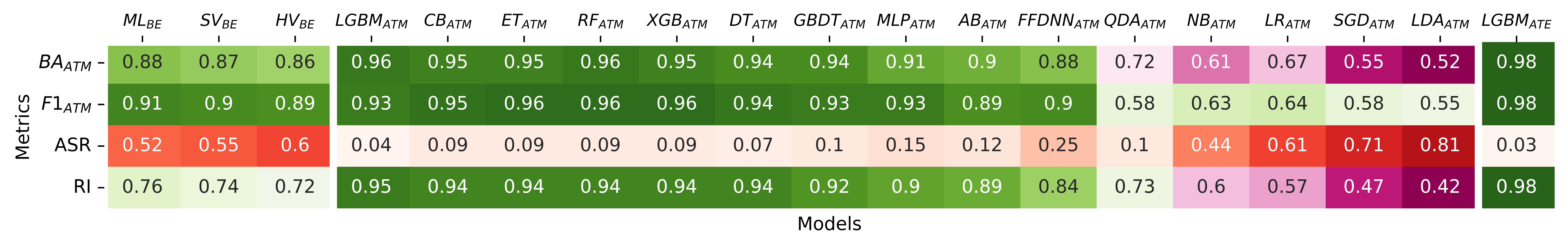}}
\caption{Robustness Index values for the evaluated defense models.}
\label{Defense Models Robustness Index}
\end{figure*}

%%%%%%%%%%%%%%%%%%%%%%%%%%%%%%%%%%%%%%%%%%%%%%%%%%%%%%%%%%%%%%%%%%%%%%%%%%%%%%%%%%%%%%
\subsection{Baseline Ensemble (BE)}
\label{Ensemble Learning}

Ensemble learning combines multiple models to improve predictive performance by aggregating different decision patterns. In adversarial settings, ensembles may improve robustness when the base models make complementary errors. However, ensemble learning does not automatically guarantee adversarial robustness, especially when transferred adversarial samples exploit weaknesses shared by several base models.

We implemented three ensemble techniques: hard voting (HV), soft voting (SV), and meta-learning (ML). In hard voting, the final prediction is based on the majority vote among the base models. In soft voting, class probabilities are averaged or weighted, and the class with the highest aggregated probability is selected. In meta-learning, a higher-level classifier is trained using the outputs or features of base models. The base models were selected from the strongest clean-data performers reported in Table~\ref{Baseline Models Performance over Clean vs Adversarial Data}, including RF, CatBoost, DT, ET, XGB, GBDT, LGBM, FF-DNN, and MLP.

As shown in Figure~\ref{Defense Models Robustness Index}, all baseline ensemble models achieve high balanced accuracy on clean data. However, their robustness on the robustness evaluation set is weaker than expected. The average balanced accuracy of the ensemble models on the robustness evaluation set is 0.87, which is lower than individual models such as LGBM and MLP. Meta-learning obtains the lowest ASR among the baseline ensemble methods, with an ASR of 0.52, followed by soft voting at 0.55 and hard voting at 0.60. Their average RI is 0.74, which remains lower than the RI values of the strongest individual baseline models, such as LGBM, MLP, and GBDT.

These results indicate that ensemble learning alone does not consistently improve robustness against transferable adversarial examples in this setting. Although ensembles are effective for clean-data performance, their adversarial robustness depends on the diversity and complementarity of the base models. When base models share similar vulnerabilities, aggregating their predictions may not reduce attack success.

%%%%%%%%%%%%%%%%%%%%%%%%%%%%%%%%%%%%%%%%%%%%%%%%%%%%%%%%%%%%%%%%%%%%%%%%%%%%%%%%%%%%%%
\subsection{Adversarially Trained Models (ATM)}
\label{Adversarially Trained Models}

Adversarial training improves robustness by exposing models to adversarial examples during training. In this study, the baseline models were retrained using clean and adversarial samples. Unlike the baseline binary evaluation, the adversarially trained models were trained as multiclass classifiers with three labels: \textit{Normal}, \textit{Attack}, and \textit{Adversarial}. This allows the models to explicitly distinguish adversarially perturbed attack samples from ordinary attack samples.

The results in Figure~\ref{Defense Models Robustness Index} show that adversarial training improves robustness compared with the baseline models. The adversarially trained models achieve an average balanced accuracy of 0.82 and an average macro F1-score of 0.82, compared with 0.79 and 0.80 for the baseline models on the robustness evaluation set. More importantly, the average ASR decreases from 0.56 for the baseline models to 0.25 for the adversarially trained models, corresponding to a reduction of 44.46\%. The average RI also increases from 0.68 for the baseline models to 0.80 for the adversarially trained models. These results suggest that adversarial training improves empirical robustness under the evaluated attack set. However, the improvement is not uniform across all attacks, as shown later in Section~\ref{Defense Effectiveness Against Adversarial Attacks}. This indicates that adversarial training is beneficial, but its effectiveness depends on the attack type and the adversarial samples used during training.

%%%%%%%%%%%%%%%%%%%%%%%%%%%%%%%%%%%%%%%%%%%%%%%%%%%%%%%%%%%%%%%%%%%%%%%%%%%%%%%%%%%%%%
\begin{figure}[!htpb]
\centering
\fbox{\includegraphics[width=0.95\columnwidth]{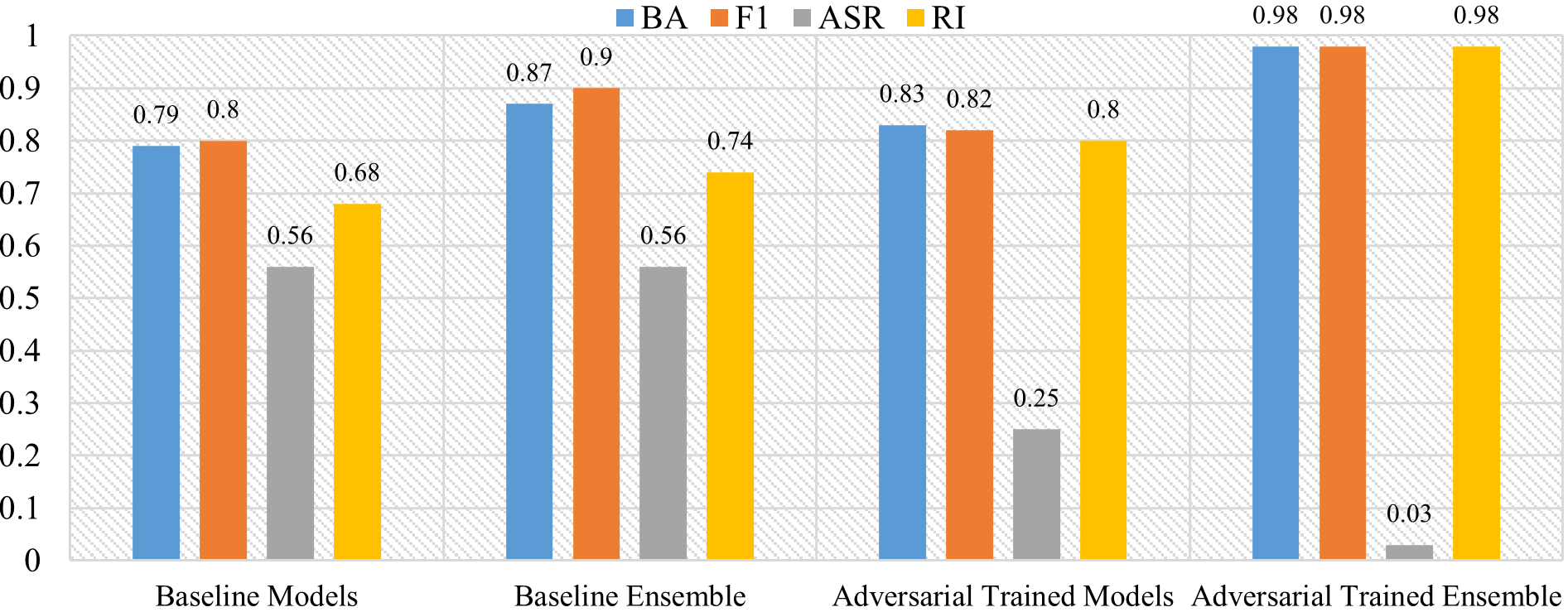}}
\caption{Comparison between baseline and defense models.}
\label{Comparison Baseline Models vs. Defense Models}
\end{figure}

%%%%%%%%%%%%%%%%%%%%%%%%%%%%%%%%%%%%%%%%%%%%%%%%%%%%%%%%%%%%%%%%%%%%%%%%%%%%%%%%%%%%%%
\subsection{Adversarially Trained Ensemble (ATE)}
\label{Adversarially Trained Ensemble}

The adversarially trained ensemble combines adversarial training with ensemble-based learning. The motivation is to use adversarial training to expose the model to perturbed samples while using ensemble learning to improve stability and reduce dependence on a single decision structure. Based on the results in Figure~\ref{Defense Models Robustness Index}, LGBM obtains the highest RI among the adversarially trained models, with an RI of 0.95. Therefore, LGBM was selected as the base learner for constructing the homogeneous ATE using bagging and boosting techniques. Bagging trains multiple models on different subsets of the training data and aggregates their predictions~\cite{buczak2015survey}. Boosting iteratively trains models by focusing on misclassified instances, producing a stronger final model~\cite{buczak2015survey}.

As shown in Figure~\ref{Defense Models Robustness Index}, the proposed ATE achieves the strongest performance among the evaluated defense configurations. It reduces ASR to 0.03, compared with 0.56 for the baseline models and 0.25 for the adversarially trained models. It also achieves balanced accuracy and macro F1-score values of 0.98. Using RI, the ATE obtains the highest score, with an RI of 0.98. In comparison, the average RI values are 0.68 for baseline models, 0.74 for baseline ensembles, and 0.80 for adversarially trained models. These results indicate that combining adversarial training with an LGBM-based ensemble provides the strongest empirical robustness among the evaluated defenses. The improvement is reflected not only in lower attack success but also in high classification performance on the robustness evaluation set. Therefore, ATE provides the most favorable trade-off between classification performance when adversarial samples are included and evasion resistance under the evaluated transfer-based setting.

%%%%%%%%%%%%%%%%%%%%%%%%%%%%%%%%%%%%%%%%%%%%%%%%%%%%%%%%%%%%%%%%%%%%%%%%%%%%%%%%
\subsection{Performance of the Adversarial Attack Classifier}
\label{Performance of Adversarial Attacks Classifier}

The second stage of AR-NIDS is the adversarial attack classifier, which identifies the type of adversarial attack associated with samples detected as adversarial. The classifier is implemented using LGBM. As shown in Table~\ref{Adversarial Attacks Classifier Performance}, it achieves an overall accuracy of 0.97, with macro-average precision, recall, and F1-score values of 0.97.

The classifier performs particularly well on DeepFool, FGSM, PGD, and ZOO, each of which achieves an F1-score of 1.00. CW2, CW\(\infty\), HSJ, and JSMA are more difficult to distinguish, with F1-scores ranging from 0.93 to 0.95. These results indicate that the generated adversarial samples contain attack-specific patterns that can be learned by the classifier, although some attacks produce more similar perturbation patterns than others. The classifier is evaluated as a closed-set classifier over the attack types generated in this study; therefore, its performance should not be interpreted as open-set detection of unseen adversarial methods.

\begin{table}[!t]
\scriptsize
\centering
\renewcommand{\arraystretch}{0.92}
\begin{tabular}{lcccc}
\toprule
\textbf{Attack} & \textbf{Precision} & \textbf{Recall} & \textbf{F1-Score} & \textbf{Support} \\
\midrule
\textbf{CW2}       & 0.94 & 0.94 & 0.94 & 25653 \\
\textbf{CW\(\infty\)} & 0.92 & 0.95 & 0.93 & 25653 \\
\textbf{DeepFool}  & 1.00 & 1.00 & 1.00 & 25653 \\
\textbf{FGSM}      & 1.00 & 1.00 & 1.00 & 25653 \\
\textbf{HSJ}       & 0.96 & 0.93 & 0.95 & 25653 \\
\textbf{JSMA}      & 0.95 & 0.94 & 0.94 & 25653 \\
\textbf{PGD}       & 1.00 & 1.00 & 1.00 & 25653 \\
\textbf{ZOO}       & 1.00 & 1.00 & 1.00 & 25653 \\
\midrule
\textbf{Accuracy}     &      &      & 0.97 & 205224 \\
\textbf{Macro Avg}    & 0.97 & 0.97 & 0.97 & 205224 \\
\textbf{Weighted Avg} & 0.97 & 0.97 & 0.97 & 205224 \\
\bottomrule
\end{tabular}
\renewcommand{\arraystretch}{1}
\caption{Performance of the adversarial attack classifier.}
\label{Adversarial Attacks Classifier Performance}
\end{table}

%%%%%%%%%%%%%%%%%%%%%%%%%%%%%%%%%%%%%%%%%%%%%%%%%%%%%%%%%%%%%%%%%%%%%%%%%%%%%%%%
\subsection{Impact of Defense Strategies on Adversarial Attack Success}
\label{Defense Effectiveness Against Adversarial Attacks}

\begin{figure}[!t]
\centering
\fbox{\includegraphics[width=0.95\columnwidth]{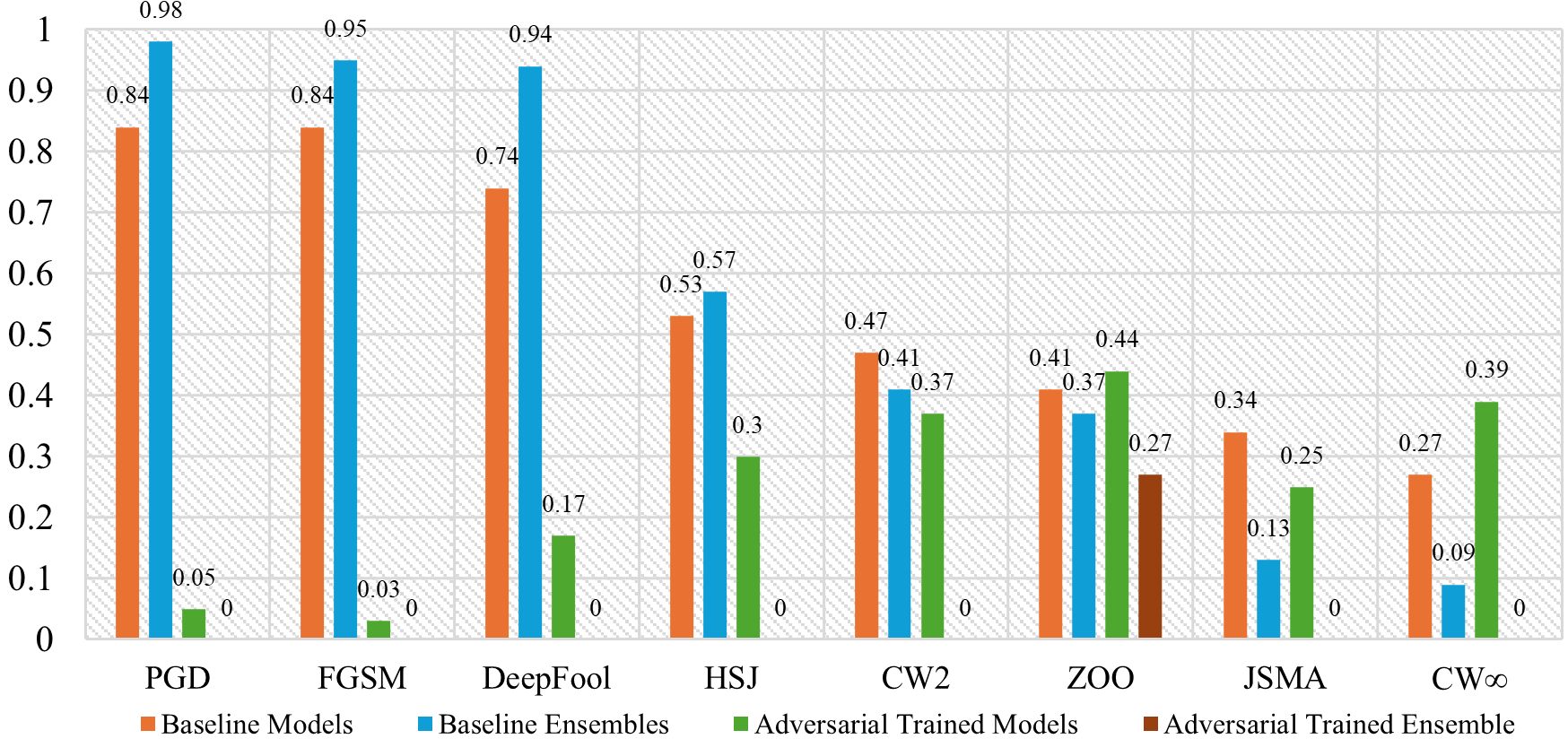}}
\caption{Impact of defense strategies on adversarial attack success.}
\label{Impact of Defenses on Adversarial Attacks}
\end{figure}

Figure~\ref{Impact of Defenses on Adversarial Attacks} compares the ASR of the evaluated attacks under different defense strategies. The results show that defense effectiveness is attack-dependent: a defense strategy that reduces the success of one group of attacks may be less effective against another group.

For PGD, FGSM, DeepFool, HSJ, and CW2, adversarial training substantially reduces attack success. For example, the ASR drops to 0.05 for PGD and 0.03 for FGSM under adversarial training. In contrast, baseline ensemble learning is less effective against these attacks, with ASR values reaching 0.98 for PGD and 0.95 for FGSM. This suggests that, in the evaluated setting, adversarial training is more effective than baseline ensemble learning for these attacks.

For ZOO, JSMA, and CW\(\infty\), the pattern is different. Baseline ensemble learning reduces the ASR of JSMA and CW\(\infty\) to 0.13 and 0.09, respectively. In contrast, adversarial training results in higher ASR values for ZOO and CW\(\infty\), reaching 0.44 and 0.39, respectively. This indicates that the effectiveness of adversarial training and ensemble learning varies across attack types.

The proposed ATE provides the most consistent reduction in attack success across the evaluated attacks. It reduces the ASR to zero for all attacks except ZOO, which is reduced to 0.27. This shows that combining adversarial training with ensemble-based learning improves robustness across the evaluated attack set more consistently than using either strategy alone.

Overall, the results demonstrate that defense effectiveness in ML-based NIDS is not uniform across attacks. Adversarial training, ensemble learning, and ATE each behave differently depending on the attack type. Among the evaluated defenses, ATE provides the strongest and most consistent empirical robustness under the transfer-based setting considered in this study.
%%%%%%%%%%%%%%%%%%%%%%%%%%%%%%%%%%%%%%%%%%%%%%%%%%%%%%%%%%%%%%%%%%%%%%%%%%%%%%%%

\section{Discussion}
\label{discussion}

This section discusses the implications of the empirical results for ML-based NIDS robustness, focusing on clean-data performance, attack transferability, defense effectiveness, and the proposed adversarially trained ensemble.

The results show that strong clean-data performance does not necessarily imply adversarial robustness. Although ensemble and tree-based models such as CatBoost, DT, ET, RF, and XGBoost perform well on clean data, several of them degrade when adversarial samples are introduced, particularly in attack recall. This indicates that clean-data accuracy or F1-score alone can overestimate the reliability of ML-based NIDS under adversarial conditions.

Attack transferability also varies across attack methods and target models. PGD, FGSM, and DeepFool show the strongest transfer effectiveness, suggesting that adversarial examples generated against the FF-DNN surrogate can exploit vulnerabilities shared by multiple detection models. In contrast, ZOO, JSMA, and CW\(\infty\) show lower or more variable transfer success. This highlights the need to evaluate robustness using multiple attack families rather than relying on a single adversarial method.

The defense results further show that robustness is attack-dependent. Baseline ensemble learning improves clean-data performance but does not consistently reduce transferred adversarial success, especially when base models share similar weaknesses. Adversarial training provides stronger robustness gains overall, reducing average attack success and improving RI, but its effectiveness still varies across attack types. These findings indicate that no single defense mechanism should be assumed robust without attack-specific evaluation.

The proposed adversarially trained ensemble (ATE) provides the strongest empirical robustness among the evaluated defenses. By combining adversarial training with an LGBM-based ensemble, ATE reduces attack success while maintaining high classification performance on the robustness evaluation set. This supports the value of combining complementary defense strategies rather than relying on ensemble learning or adversarial training alone. However, the result should be interpreted within the evaluated transfer-based setting and attack set, not as a universal robustness guarantee.

The adversarial attack classifier complements AR-NIDS by identifying the type of adversarial manipulation associated with detected adversarial samples. Its high accuracy suggests that the evaluated attack methods leave distinguishable patterns in the generated adversarial samples. However, because the classifier is evaluated on the same attack families considered in this study, identifying unseen adversarial methods remains an important direction for future work.

Overall, the findings show that adversarial robustness in ML-based NIDS depends on the interaction between attack type, model architecture, and defense strategy. The proposed RI helps summarize these trade-offs by combining balanced accuracy and macro-F1 score computed on the robustness evaluation set with ASR into a compact comparative score.

\textbf{Scope and Future Work.}
The findings should be interpreted within the evaluated transfer-based setting and attack set. The study uses the NF-UQ-NIDS unified NetFlow-based benchmark, which provides diverse traditional and IoT traffic with multiple attack categories; however, validation on additional datasets and feature representations would strengthen generalizability. In addition, adversarial examples are generated over feature representations, making the study suitable for controlled assessment of model sensitivity and transferability. Future work will examine additional attack configurations, protocol-constrained adversarial generation, and broader validation of the adversarial attack classifier against unseen attack types.

%============================================================
\section{Conclusion}\label{conclusion}

This paper evaluated ML-based NIDS robustness under transferable adversarial attacks using eight evasion attacks, fifteen detection models, and three representative defense strategies on the NF-UQ-NIDS dataset. The results show that strong clean-data performance does not necessarily imply adversarial robustness, as several models degraded when adversarial samples were introduced. The attack-level analysis further showed that PGD, FGSM, and DeepFool were the most effective transferable attacks, while other attacks showed more variable success across models.

To support comparative robustness assessment, we introduced the Robustness Index (RI), which combines balanced accuracy and macro-F1 score computed on the robustness evaluation set with attack success rate. RI showed that robustness varies across model architectures, with LGBM, MLP, and GBDT achieving stronger baseline robustness under the evaluated transfer-based setting.

The proposed AR-NIDS framework combines an adversarially trained ensemble with an adversarial attack classifier. The adversarially trained ensemble achieved the strongest trade-off among the evaluated defenses, reducing the average attack success rate from 0.41 to 0.03 and achieving an RI of 0.98, while the adversarial attack classifier achieved 0.97 accuracy in identifying the evaluated attack types. These findings support the use of combined adversarial training and ensemble-based modeling for improving empirical robustness in ML-based NIDS.

%%%%%%%%%%%%%%%%%%%%%%%%%%%%%%%%%%%%%%%%%%%%%%%%%%%%%%%%

%%%%%%%%%%%%%%%%%%%%%%%%%%%%%%%%%%%%%%%%%%%%%%%%%%%%%%%%
\FloatBarrier
% -----------------------------------------------------------------------------
% Layout compression log (content preserved):
% - Reduced float/caption/list/display spacing in the preamble.
% - Slightly reduced widths of major figures to save vertical space.
% - Compacted large tables using smaller table fonts, reduced tabcolsep, and reduced arraystretch.
% - Compacted bibliography spacing.
% -----------------------------------------------------------------------------

\bibliographystyle{elsarticle-num}
\scriptsize
\bibliography{cas-refs}

\end{document}